\documentclass[hyper,a4paper]{JHEP3}

\usepackage{epsfig}
\usepackage{latexsym}
\usepackage{graphicx}
\usepackage{amsmath}
\usepackage{amsfonts}
\usepackage{amssymb}
\usepackage{float}
\usepackage{bm}
\usepackage{cite}
\newcommand{\mathsym}[1]{{}}

\newcommand{\baz}{\begin{array}{cc}}
\newcommand{\bad}{\begin{array}{ccc}}
\newcommand{\bi}{\begin{itemize}}
\newcommand{\ei}{\end{itemize}}
\newcommand{\ba}{\begin{array}{c}}
\newcommand{\ea}{\end{array}}

\def\be{\begin{equation}}
\def\ee{\end{equation}}
\def\bea{\begin{eqnarray}}
\def\eea{\end{eqnarray}}\def\nn{\nonumber}
\def\gsim{\ \rlap{\raise 2pt\hbox{$>$}}{\lower 2pt \hbox{$\sim$}}\ }
\def\lsim{\ \rlap{\raise 2pt\hbox{$<$}}{\lower 2pt \hbox{$\sim$}}\ }
\def\dslash{\kern-4pt \not{\hbox{\kern-2pt $\partial$}}}
\def\pslash{\not{\hbox{\kern-2pt p}}}
\def\pmue{{${\rm P_{\mu e}}$ }}
\def\pmumu{{${\rm P_{\mu \mu}}$ }}

\def\dam{\mathrm{(\Delta m_{31}^2)^m}}

\def\da{\mathrm{\Delta m_{31}^2}}

\def\dcp{{\delta_{\mathrm{CP}}}}
\def\stsmallm{\mathrm{\sin^2 2 \theta_{13}^m}}
\def\peem{{{\rm P^m_{e e}}}}
\def\pmue{{{\rm P_{\mu e}}}}
\def\pmuem{{{\rm P^m_{\mu e}}}}
\def\pemum{{{\rm P^m_{e \mu}}}}
\def\pemu{{{\rm P_{e \mu}}}}
\def\pmumu{{{\rm P_{\mu \mu}}}}
\def\pmumum{{{\rm P^m_{\mu \mu}}}}
\newcommand{\nova}{NO$\nu$A}
\newcommand{\dmatm}{\Delta m_{atm}^2 }
\newcommand{\tmm}{\theta_{\mu\mu}}

\DeclareGraphicsExtensions{.eps,.ps}
\numberwithin{equation}{section}

\title{Octant sensitivity for large $\theta_{13}$ in 
atmospheric and long baseline neutrino experiments }

\author{Animesh Chatterjee$^a$, Pomita Ghoshal$^b$, Srubabati Goswami$^b$, 
Sushant K. Raut$^b$ \\
\llap{$^a$} Harish-Chandra Research Institute, Chhatnag Road, Jhunsi,
Allahabad 211 019, India \\
\llap{$^b$} Physical Research Laboratory, Navrangpura,
Ahmedabad 380 009, India \\
E-mail: \email{animesh@hri.res.in}, \email{pomita@prl.res.in}, \email{sruba@prl.res.in}, 
\email{sushant@prl.res.in} }

\date{\today}
\abstract{
 One of the unknown parameters in neutrino oscillation studies is 
 the octant of the atmospheric neutrino mixing angle $\theta_{23}$. 
 In this paper, we discuss the possibility of determining 
 the octant  of $\theta_{23}$ in the long baseline experiments T2K and 
 \nova\ in conjunction with future atmospheric neutrino detectors, 
 in the light of  non-zero value of $\theta_{13}$  measured by reactor experiments. 
 We consider two detector technologies for atmospheric neutrinos --  
 magnetized iron calorimeter and  non-magnetized Liquid Argon Time Projection Chamber. 
 We present the octant sensitivity for T2K/\nova\ and 
 atmospheric neutrino experiments separately as well as the combined sensitivity. 
 For the long baseline experiments, a precise measurement of 
 $\theta_{13}$, which can exclude degenerate solutions in the wrong octant,
 increases the sensitivity drastically.  
 For $\theta_{23} = 39^o$ and $\sin^2 2\theta_{13}=0.1$, 
 at least $\sim 2\sigma$ sensitivity can be achieved by T2K + \nova\
 for all values of $\delta_{CP}$ for both normal and inverted hierarchy. 
 For atmospheric neutrinos, the moderately large value of $\theta_{13}$ 
 measured in the reactor experiments is conducive to octant sensitivity
 because of enhanced matter effects. A magnetized iron detector can give a  
 2$\sigma$ octant sensitivity  for 500 kT yr exposure for 
 $\theta_{23} = 39^o$, $\delta_{CP} =0$  and normal hierarchy. 
 This increases to $3\sigma$ for both hierarchies by combining with T2K and \nova. 
 This is due to 
 a preference of different $\theta_{23}$ values at the minimum $\chi^2$ 
 by T2K/\nova\ and atmospheric neutrino  experiments. 
 A Liquid Argon type detector for atmospheric neutrinos with the same exposure 
 can give higher octant sensitivity, 
 due to the interplay of 
 muon and electron contributions and superior resolutions. 
 We obtain a $\sim 3\sigma$ sensitivity for $\theta_{23} = 39^o$ for normal hierarchy.  
 This increases to $\gsim 4\sigma$ for all values of  $\delta_{CP}$ 
 if combined with T2K/\nova. For inverted hierarchy the combined sensitivity is around $3\sigma$.  
} 
\begin{document} 

\section{Introduction}

The measurement of a non-zero value of the mixing angle $\theta_{13}$ 
by the reactor experiments Double-Chooz~\cite{DC}, Daya-Bay~\cite{daya} and  
RENO~\cite{reno}  
heralds a major breakthrough in the advancement of 
neutrino  physics. 
The best-fit value of $\sin^2\theta_{13}$ from latest global analysis
of solar, atmospheric, reactor and accelerator 
data is   $0.023 \pm 0.0023$~\cite{msg_global}. 
which signifies non-zero $\theta_{13}$ at 10$\sigma$  level.  
Other 
global analyses also give similar results~\cite{lisi-global,valle-global}.
This confirms the earlier observation of non-zero $\theta_{13}$ 
in T2K~\cite{t2k}  and MINOS~\cite{minos} 
experiments as well the indication of non-zero best-fit
value of $\theta_{13}$ from  
previous analyses 
\cite{lisith13,sg-smirnov}. 
Among the other oscillation parameters the solar parameters 
$\Delta m^2_{21}$ and $\theta_{12}$ are already well measured 
from solar neutrino and KamLAND experiments~\cite{sno, kamland}.  
The solar matter effect 
also dictates $\Delta m^2_{21} >0$.  The most stringent 
constraint on  mass squared difference 
$\Delta m^2_{31}$ governing the atmospheric neutrino oscillations 
comes from the data from the MINOS experiment~\cite{minos_2012}. 
However the ordering of the third mass eigenstate with respect to
the other two {\it i.e.} the sign of $\Delta m^2_{31}$ is not yet known. 
There are two possible arrangements of the neutrino mass states: 
(i) $m_1 < m_2 < m_3$ corresponding to Normal Hierarchy (NH)
and (ii) $m_3 < m_1 < m_2$ corresponding
to Inverted Hierarchy (IH).  
The 
mixing angle $\sin^2\theta_{23}$ is mainly determined by 
the SuperKamiokande (SK) atmospheric neutrino data. 
However the octant 
in which this mixing angle lies is not yet decisively 
determined by the data. 
A full three-flavour fit to the SK data gives the best-fit 
for NH in the lower octant (LO) and IH in the higher octant (HO) 
keeping $\theta_{13}$ as a free parameter in the analysis~\cite{sk}. 
Three-flavour global analysis of all available neutrino data 
give the best-fit $\theta_{23}$ in the lower octant. 
The best-fit values and 3$\sigma$ ranges of oscillation parameters 
from global analysis in Ref.~\cite{lisi-global} is summarized in Table~\ref{table-osc}. 
At present there is no significant constraint on the CP phase $\delta_{CP}$
and the whole range from $0- 2\pi$ is allowed at the 2$\sigma$ level.

\begin{table}[t]\centering
  \begin{tabular}{|c|c|c|}
    \hline
    parameter & best fit & 3$\sigma$ range
    \\
    \hline
    $\Delta m^2_{21}\: [10^{-5}~\text{eV}^2]$
    & 7.54 & 6.99--8.18 \\[1.5mm] %%
    $|\Delta m^2_{31}|\: [10^{-3}~\text{eV}^2]$
    &
    \begin{tabular}{c}
      2.43\\
      2.42
    \end{tabular}
    &
    \begin{tabular}{c}
      $2.19-2.62$\\
      $2.17-2.61$
    \end{tabular}
    \\[4.5mm]
 $\sin^2\theta_{12}$
    & 0.307 & 0.26--0.36\\[2.5mm]  %%
    $\sin^2\theta_{23}$
    &
    \begin{tabular}{c}
      0.386\\
      0.392
    \end{tabular}
&
\begin{tabular}{c}
      0.33--0.64\\
      0.34--0.66
    \end{tabular}
    \\[4mm]
    $\sin^2\theta_{13}$
    &
    \begin{tabular}{c}
      0.0241\\
      0.0244
    \end{tabular}
    &
    0.017--0.031 
   \\[4mm]
    $\delta$
   &
   \begin{tabular}{c}
     $1.08\pi$\\
     $1.09\pi$
   \end{tabular}
   &
   $0-2\pi$ \\
       \hline
     \end{tabular}
\caption{ \label{table-osc} The best-fit values and $3\sigma$ ranges of neutrino
       oscillation parameters from Ref.~\cite{lisi-global }. The first and second rows
in each case denote the values for a normal and inverted neutrino
mass hierarchy respectively. 
       }
\end{table}

%However 
%the measurement of a moderately large value of $\theta_{13}$ 
%is likely to pave the way towards the discovery of  the three 
%^unknowns  discussed above --- $\delta_{CP}$  
%mass hierarchy and octant of $\theta_{23}$ and charter the future 
%course in neutrino oscillation experiments.  
%A large value of $\theta_{13}$  also signifies 
%sizeable earth matter effect during neutrino propagation.
There are two aspects in the measurement 
of $\theta_{13}$ which is expected to play an important role  
in resolving the outstanding issues in neutrino oscillation physics. 
Firstly if  $\theta_{13}$ is     
non-zero and relatively large  --- 
(i)  CP violation in the lepton sector  can be probed~~
(ii) the earth matter effect on the propagation of 
neutrinos can be sizable. 
The latter facilitates the determination of 
mass hierarchy and octant in experiments in which 
neutrinos travel through an appreciable path length. 
The second aspect is the precision measurement of $\theta_{13}$: 
which helps in increasing the sensitivity in the determination of 
hierarchy, octant and $\delta_{CP}$.

The octant degeneracy means impossibility of 
distinguishing between $\theta_{23}$ and $\pi/2 -\theta_{23}$. 
This is generic and robust for  vacuum oscillation 
probabilities that are functions 
of $\sin^2 2\theta_{23}$ e.g. the two flavour muon survival probability 
in vacuum~\cite{lisioctant}. If on the other hand the leading term 
in the probability are functions of 
$\sin^2\theta_{23}$  (e.g. $P_{\mu e}$) then the inherent 
octant degeneracy is not there but lack of knowledge of other 
parameters like $\theta_{13}$ and  $\delta_{CP}$  can give rise to  
octant degeneracy~\cite{degen_8fold,ourpaper1}. 
%It had been realized earlier that a precise determination of 
%$\theta_{13}$ by the reactor experiments can alleviate this problem 
%\cite{lindner,minakata}. However one can still ask how much the lack 
%of knowledge of $\delta_{CP}$ 
%and the precise value of $\theta_{23}$ in the wrong 
%can affect this degeneracy.  
These issues may affect the octant sensitivity of the long baseline
experiments  
T2K and \nova\ where the matter effect is not very significant
and in particular resonant matter effects do not get a chance to 
develop. 
Although conventionally the  octant degeneracy
refers to the indistinguishability of $\theta_{23}$ and $\pi/2 -\theta_{23}$,
in view of the present uncertainty in the measurement of $\theta_{23}$
the scope of this can be generalized to  any value of  $\theta_{23}$
in the wrong octant within its allowed range.
In this paper by octant sensitivity we refer to this ``generalized"
definition.
If in addition the hierarchy is unknown then 
there can also be the wrong-octant--wrong-hierarchy solutions and 
one  needs to  marginalize 
over the wrong hierarchy  as well.

% for baselines relevant for long baseline 
%experiments as well as  atmospheric neutrinos.  
%The atmospheric neutrinos of course 
%In the two flavour approximation  ($\theta_{13}=0$) 
%the muon  neutrino survival probability ($P_{\mu \mu}$)  
%in vacuum is a function of $\sin^2 2 \theta_{23}$ and hence cannot discriminate 
%between $\theta_{23}$ and $\pi/2 - \theta_{23}$. 
%In presence of non-zero $\theta_{13}$ the $P_{\mu e}$ channel also 
%contributes and the leading term contains $\sin^2\theta_{23} \sin^2 2\theta_{13}$. However although $\sin^2\theta_{23}$ does not suffer from octant degeneracy 
%in this case, the degeneracy due to $\theta_{13}$ can come in. 
% it is possible that the above combination remains the 
%same in opposite octants for different values of $\theta_{13}$. 
Atmospheric neutrinos 
pass through long distances in matter and they span a wide 
range in energy and can  
encounter resonant matter effects. In this case
the octant sensitivity in 
$P_{\mu \mu}$ ensues from the term $\sin^4 \theta_{13} \sin^2 2\theta_{13}^m$ 
\cite{choubey-roy}. $P_{\mu e}$ in matter contains
$\sin^2 \theta_{13} \sin^2 2\theta_{13}^m$. 
Since at resonance $\sin^2 2\theta_{13}^m \approx 1$, the octant 
degeneracy can be removed. 
In this case also one can 
probe the effect of $\delta_{CP}$ uncertainty
on the lifting of this degeneracy. 

Atmospheric neutrinos provide fluxes of both neutrinos and antineutrinos 
as well as neutrinos of both electron and muon flavour.  
On one hand it provides the advantage of observing both electron and muon 
events. However on the other hand a particular type of event gets contributions 
from both disappearance and appearance probabilities. This can be a problem 
if the matter effects for these two channels go in opposite directions. 
Thus it is necessary to carefully study the various contributions and ascertain 
what may be the best possibility to decode the imprint of matter effects in 
atmospheric neutrino propagation. 
Three major types of detector technologies are under consideration 
at the present moment as future detector of atmospheric neutrinos. 
\\ 
(i) Water Cerenkov detectors : Such type of detectors have already been shown 
to be a successful option for atmospheric neutrino detection by the 
SuperKamiokande 
collaboration. This is sensitive to both electron and muon events and 
the energy threshold  can be relatively low. 
Megaton detectors of this kind under consideration for future are 
HK, MEMPHYS~\cite{hkloi,memphys}. These cannot be magnetized and hence provide no 
charge identification capability. 
% The octant discrimination capability of HK has been discussed in Ref.~\cite{hkloi}. 
Multi-megaton detectors with ice also fall in this category.  
An example of such a detector is PINGU~\cite{pingu,pingu2}, which 
is a proposed upgrade of the DeepCore section of the IceCube detector for ultra-high 
energy neutrinos~\cite{icecube_deepcore}
with a lower energy threshold for atmospheric neutrino detection. 
\\
(ii) Magnetized Iron Detectors: 
Such a detector for atmospheric neutrinos were proposed by the MONOLITH
\cite{monolith} 
collaboration and is now actively pursued by the INO collaboration~\cite{ino}. 
This has a relatively high  threshold 
and is mainly sensitive to muon neutrinos. 
%\footnote{For sensitivity to 
%electron neutrinos one needs thinner iron plates which can 
%increase background and   also costlier} 
These type of detectors offer the possibility of magnetization, 
thus making it possible to distinguish 
between muon and antimuon events, which enhances the sensitivity. 
\\ 
(iii) Liquid Argon Time Projection Chamber (LArTPC):  
Examples of such detectors are ICARUS and ArgoNeuT~\cite{icarus, argoneut}. 
The hallmark of these detectors are their 
superior particle identification capability and 
excellent energy and angular resolution. 
They are sensitive to both electron and muon events 
with good energy and direction reconstruction capacity for 
both type of events~\cite{larmag0}. 
Possibility of magnetization is also being discussed~\cite{larmag1,larmag2}.

In this paper we study in detail the possibility of removal 
of octant degeneracy in view of the precise measurement of a 
relatively large value of $\theta_{13}$. 
There have been many earlier studies dealing with this subject
both in the context of long baseline and atmospheric neutrinos.
The importance of combining accelerator and reactor 
experiments and the role of a 
precise measurement of $\theta_{13}$ in 
abating 
the octant degeneracy have been considered 
in Refs.~\cite{lindner,minakata,minakata2}.  
The possibility of resolving the octant degeneracy using 
long baseline experiments has also been explored in 
Refs.~\cite{octant_lbl,octant_matter_lbl,dev_ggms,octant_lbl_3flav,octant_t2kk,octant_silver}. 
Recently octant sensitivity in the  
T2K/\nova\ experiments has been investigated including the recent 
results on the measurement of non-zero $\theta_{13}$ by reactor 
experiments~\cite{uss}. The octant sensitivity for atmospheric 
neutrinos in the context of magnetized iron calorimeter 
detectors was considered in Refs.~\cite{choubey-roy,nita_th23,samanta}, for 
water Cerenkov detectors in Ref.~\cite{hkloi} and for 
LArTPC in Ref.~\cite{ourprl}. 
 
We examine the octant sensitivity in the  
long baseline experiments T2K and \nova\  
and  in the atmospheric neutrino experiments as well 
as the combined sensitivity of these experiments. 
In particular we address whether degeneracy due to 
$\theta_{13}$ can still affect octant determination
at the current level of precision 
of this parameter.  For fixed values of 
$\theta_{13}$ the effect of lack of knowledge of $\delta_{CP}$ 
on the octant determination capability of these 
experiments is also studied.
We take into account the uncertainty of $\theta_{23}$ 
in the wrong octant and  discuss how much this can influence the octant 
sensitivity. 
We present results for the two cases of known and 
unknown hierarchy. 
 
For the study of atmospheric neutrinos 
we consider magnetized iron calorimeter detectors
with charge sensitivity which is sensitive to the muon neutrinos. 
We also consider a non-magnetized LArTPC detector which can 
detect both electron and muon neutrinos. 
In particular we discuss the interplay between the muon and electron 
type events in the overall octant sensitivity. 
For our atmospheric analysis we assume a prior knowledge of hierarchy. 
Lastly we do a combined analysis of T2K, \nova\ and atmospheric 
neutrinos and discuss the synergistic aspects between long baseline 
and atmospheric neutrino experiments. 

%Preliminary results on  octant sensitivity 
% 
%have been discussed in \cite{barger} in the context of a magnetized 
%Liquid Argon detector.  In this paper we study this 
%more extensively and study the dependence on octant sensitivity on various 
%factors like CP phase, magnetization of the detector etc.  
The plan of the paper is as follows.
In Section~\ref{sec:physics} we discuss the octant degeneracy 
at the level
of oscillation and survival probabilities, 
for baselines corresponding to  both atmospheric neutrinos
and those relevant to \nova\ and T2K. 
Section~\ref{sec:analysis} discusses the analysis procedure and results.
First we discuss the octant sensitivity in 
\nova\ and T2K. 
Next we describe the results obtained for  
octant sensitivity using atmospheric neutrino 
detectors. Finally we present combined octant sensitivity of
long baseline and 
atmospheric
neutrino experiments. 
We end by summarizing the results.

%-----------------------------------------------------------%

\section{Analysis of octant degeneracy} 
\label{sec:physics}

The ambiguity in the determination of the octant of 
$\theta_{23}$ may appear in the oscillation and 
survival probabilities as 

(a) the intrinsic octant degeneracy,
in which the probability is a function of $\sin^2 2\theta_{23}$ and
hence the measurement cannot distinguish between 
$\theta_{23}$ and $\pi/2 - \theta_{23}$,
\begin{equation} 
P(\theta_{23}^{tr}) = P(\pi/2 - \theta_{23}^{tr}) 
\end{equation} 

(b) the degeneracy of the octant with other neutrino parameters, 
which confuses octant determination due to the 
uncertainty in these  parameters.
In particular, this degeneracy arises 
in probabilities that are functions of $\sin^2\theta_{23}$ or 
$\cos^2\theta_{23}$. For such cases 
$P(\theta_{23}^{tr}) \neq P(\pi/2 - \theta_{23}^{tr})$.
However for different 
values of 
the parameters $\theta_{13}$ and $\delta_{CP}$ the probability functions 
become  identical for values of $\theta_{23}$ in 
opposite octants for different values of these 
parameters, i.e. 
\begin{equation} 
P(\theta_{23}^{tr},\theta_{13},\delta_{CP})
= P(\pi/2 - \theta_{23}^{tr},\theta'_{13},\delta'_{CP}),
\end{equation} 
where $\theta_{23}^{tr}$ denotes the true value of 
the mixing angle  and 
the primed and unprimed values of $\theta_{13}$ and  
$\delta_{CP}$ lie within
the current allowed ranges of these parameters. In the case of $\delta_{CP}$, 
this covers the entire range from $0$ to $2\pi$, while 
for $\theta_{13}$ the current 3$\sigma$ range is given by
$\sin^2 2\theta_{13} = 0.07 - 0.13$ .     
From the above equation it is evident that 
even if $\theta_{13}$ is determined
very precisely, this 
degeneracy can still remain due to complete uncertainty 
in the CP phase. 
In fact, the scope of this degeneracy can be enlarged to define this 
as 
\begin{equation} 
P(\theta_{23}^{tr},\theta_{13},\delta_{CP})
= P(\theta_{23}^{wrong},\theta'_{13},\delta'_{CP})
\end{equation}
where $\theta_{23}^{wrong}$ denote values of 
the mixing angle in the opposite octant.

The features of the octant degeneracy and the potential for its resolution
in different neutrino energy and 
baseline ranges can be understood from the expressions for the 
oscillation and survival probabilities relevant 
to specific ranges. We discuss below the probability expressions 
in the context of the fixed baseline experiments 
\nova/T2K and for atmospheric neutrino experiments.

\subsection{Neutrino Propagation in Matter} 

Neutrinos travelling through earth 
encounter a potential due to matter given as, 
\be
{\mathrm{A = 2\;\sqrt{2}\;G_F\;n_e\;E = 2 \times 0.76 \times 10^{-4} \times Y_e
\;\left[\frac{\rho}{g/cc}\right]
\;\left[\frac{E}{GeV}\right]\;eV^2 }}
\label{A_mat}
\ee
where $G_F$ is the 
Fermi coupling constant and $n_e$ is the electron number density in matter, 
given by $n_E  = N_A Y_e \rho$ ($N_A=$ 
Avogadro's number, $Y_e=$ electron fraction $\sim 0.5$, 
$\rho=$ earth matter density.  

The  mass squared difference ${\mathrm{\dam}}$ and mixing angle
${\mathrm {\stsmallm}}$ in matter are related to their vacuum values by
%-----------------------------------------------------------%
%-----------------------------------------------------------%
\bea
{\mathrm{\dam}} &=&
{\mathrm{
\sqrt{(\da \cos 2 \theta_{13} - A)^2 +
(\da \sin 2 \theta_{13})^2} }}
\nn \\
\nn \\
{\mathrm {\sin 2 \theta^m_{13} }}
&=& \frac{\mathrm{\da \sin 2 \theta_{13}}}
{{\mathrm{\sqrt{(\da \cos 2 \theta_{13} - A)^2 +(\da \sin 2 \theta_{13})^2} }}}
%&=& \frac{\Delta m^2_{31} \sin 2 \theta_{13}}
%{{\mathrm{
%{(\Delta m_{31}^2)^m} }}}
\label{eq:dm31}
\eea 

The MSW matter resonance~\cite{msw1,msw2,msw3} occurs and the mixing angle 
$\theta_{13}^{\rm{m}}$ becomes 
maximal at neutrino energies and baselines for which the terms $\da \cos 2 \theta_{13}$ 
and A in the denominator 
of eq.(\ref{eq:dm31}) become equal. Hence the matter resonance energy $E_{res}$ 
is given by
\be
{\rm{E_{res} = \frac{|\da| \cos 2\theta_{13}}{ 2 \times 0.76 \times 10^{-4} \times Y_e \times \rho}}}
\label{E_res}
\ee

Since the corresponding expression for antineutrinos is obtained by making the 
replacement $A \to -A$, it may 
be observed that the matter resonance
occurs for the normal neutrino mass hierarchy (i.e. $\da > 0$) for neutrinos 
and for the inverted mass 
hierarchy (i.e. $\da < 0$) for antineutrinos.

\subsection{Octant ambiguity in $P_{\mu e}$ and $P_{\mu \mu}$} 
\label{subsec:prob}

For \nova/T2K, the baselines are 812 and 295 Km respectively and the peak 
energies of the beams are 
in the range 0.5-2 GeV. 
For these values of baselines, the earth matter density is in the range 
$2.3 - 2.5$ g/cc, and the 
corresponding matter resonance 
energies are above 10 GeV. Hence the neutrino energies of both 
experiments lie well below matter resonance,
and the oscillation probabilities will only display small sub-leading matter 
effects. The expressions of the relevant probabilities $\pmumu$ and 
$\pmue$ in vacuum are given by the following  expressions obtained in the 
one-mass scale dominant
(OMSD) approximation, 

\begin{equation}
{\mathrm{P^{v}_{\mu \mu}}} =
{\mathrm{
1 -  \sin^2 2 \theta_{23}\; \sin^2\left[1.27 ~\da~\frac{L}{E} \right] + 4 \sin^2 \theta_{13} \; 
\sin^2 \theta_{23} \; \cos 2 \theta_{23} \; \sin^2\left[1.27 ~\da ~\frac{L}{E} \right]
}}
\label{eq:pmmu}
\end{equation}

\begin{equation}
{\mathrm{P^{v}_{\mu e}}} =
{\mathrm{
\sin^2 \theta_{23} \; \sin^2 2 \theta_{13} \; \sin^2\left[1.27 ~\da ~\frac{L}{E} \right]
}}
\label{eq:pemu}
\end{equation}

The above probability expressions have sub-leading 
corrections corresponding to small 
matter effect terms and the solar mass-squared difference 
\cite{BurguetCastell:2001ez,akhmedov,cervera,freund,kimura}. 
We observe the following 
salient features from these expressions:

(a) The disappearance channel $\pmumu$ has in its leading order a dependence on 
$\sin^2 2\theta_{23}$, and hence is dominated by the 
intrinsic octant degeneracy.
There is a small $\theta_{13}$-dependent correction in the measurement of 
$\theta_{23}$ 
which gives a minor ($\sim 1\%$) resolution of the degeneracy
if $\theta_{13}$ is known precisely.
% but this effect
%is not statistically significant.  

(b) The appearance channel $\pmue$ has the combination of parameters 
$\sin^2 \theta_{23} \sin^2 2 \theta_{13}$, and hence does not suffer from 
the intrinsic octant degeneracy. However, the degeneracy of the octant with 
the parameter $\theta_{13}$ 
comes into play, since the above combination may be invariant for opposite
octants for 
different values of $\theta_{13}$, and hence this degeneracy cannot get 
lifted with a 
measurement from such experiments alone~\cite{minakata}.
This channel can be also affected 
by the large uncertainty in  $\delta_{CP}$ when sub-leading corrections 
are included. 

%\item 

%\subsection{Atmospheric Neutrinos} 

For atmospheric neutrinos, the relevant baselines and energies are in the range 
1000 - 12500 Km and 1 - 10 GeV respectively. A large region in this L and E space
exhibits strong resonant earth matter effects, since the earth densities in 
this 
baseline range (3 - 8 g/cc) correspond to resonance energies
${\rm{E_{res} = 4 - 9}}$ GeV. Hence the relevant probability expressions $\pmuem$, 
$\peem$ and $\pmumum$ can be written, in the OMSD
approximation and with full matter effects, as~\cite{gandhi,akhmedov} 

\begin{equation}
{\mathrm{P^{m}_{\mu e}}} =
{\mathrm{
\sin^2 \theta_{23} \; \sin^2 2 \theta^m_{13} \; \sin^2\left[1.27 ~\dam ~\frac{L}{E} \right]
}}
\label{eq:pemumat}
\end{equation}

\bea
%\begin{align}
{\mathrm{P^{m}_{\mu \mu} }} &=&
%1 - {\mathrm {
%P^{m}_{\mu \tau}}} - {\mathrm {
%P^{m}_{\mu e}}}
%\nonumber \\
%&=&
{\mathrm{1 - \cos^2 \theta^m_{13} \; {\mathrm{\sin^2 2 \theta_{23}}}
\;
\sin^2\left[1.27 \;\left(\frac{\da + A + \dam}{2}\right) \;\frac{L}{E} \right]}}
\nonumber \\
&& ~-~
{\mathrm{
\sin^2 \theta^m_{13}\; {\mathrm{\sin^2 2 \theta_{23}}}
\;
\sin^2\left[1.27 \;\left(\frac{\da + A - \dam}{2}\right) \;\frac{L}{E}
\right]}}
\nonumber \\
&& ~-~
{\mathrm{\sin^4 \theta_{23}}} \;
{\mathrm { \sin^2 2\theta^m_{13} \;
\sin^2 \left[1.27\; \dam  \;\frac{L}{E} \right]
%\left(1.27 (\Delta m_{31}^2)^m {\mathrm L}/{\mathrm E} \right)
}}
\label{eq:pmumumat1}
\eea

\begin{equation}
{\mathrm{P^{m}_{e e}}} =
{\mathrm{
1 - \sin^2 2 \theta^m_{13}\; \sin^2\left[1.27 ~\dam~\frac{L}{E} \right]
}}
\label{eq:peemat}
\end{equation}

%The anti-neutrino probabilities are obtained by making the replacement A 
%$\to$ -A in the above expressions.
In this case, the following features are observed:

(a) The oscillation probability in matter $\pmuem$ is still guided to leading 
order by a dependence on $\sin^2 \theta_{23}$. 
%, which could suffer from a 
%degeneracy with $\theta_{13}$ and/or $\delta_{CP}$ (when sub-leading corrections 
%are included)  if there is a large enough 
%uncertainty in the latter parameters.
But strong resonant earth matter effects help 
in resolving 
the degeneracy since the mixing angle $\theta_{13}$ in matter gets amplified 
to maximal values (close to $45^o$) near resonance. The combination 
$\sin^2 \theta_{23} \sin^2 2 \theta_{13}^{\rm{m}}$ no longer remains 
invariant over opposite octants, since $\sin^2 2 \theta_{13}^{\rm{m}}$ becomes 
close to 1 in both octants irrespective of the vacuum value of $\theta_{13}$. 
This breaks the degeneracy of the octant with $\theta_{13}$.
% and $\dcp$. 

(b) The muon survival probability in matter $\pmumum$ has leading terms 
proportional to $\sin^2 2\theta_{23}$, as in the vacuum case, 
which could give rise to the intrinsic octant degeneracy. But  
the strong octant-sensitive behaviour of 
the term $\sin^4 \theta_{23} \sin^2 2 \theta_{13}^{\rm{m}}$ near resonance can 
override the degeneracy present in the 
$\sin^2 2\theta_{23}$-dependent terms.

(c) The electron survival probability $\peem$ is independent of $\theta_{23}$ 
and hence does not contribute to the octant sensitivity.

(d) Since the $\pemum$ channel is simply the CP conjugate of $\pmuem$, the 
probability level discussion in this section is applicable to $\pemum$ also.

In the following discussion, 
we address the octant degeneracy due to $\theta_{23}$ in the wrong octant, 
$\theta_{13}$ and unknown values of $\delta_{CP}$, at a 
probability level. 
The probability figures~\ref{Pvsth13_delcptest} -~\ref{PvsE_th23delcptest} 
are drawn by solving the full three flavour 
propagation equation 
of the neutrinos in matter using PREM density profile~\cite{prem}.  
In all these figures 
the left panels
are for the \nova\ peak energy and baseline (2 GeV, 812 Km), while the
right panels are for a typical atmospheric neutrino energy and baseline (6 GeV, 5000 Km).
The top row denotes the appearance channel $P_{\mu e}$,
while the bottom row denotes the disappearance channel $P_{\mu\mu}$.

\FIGURE{ 
\epsfig{file=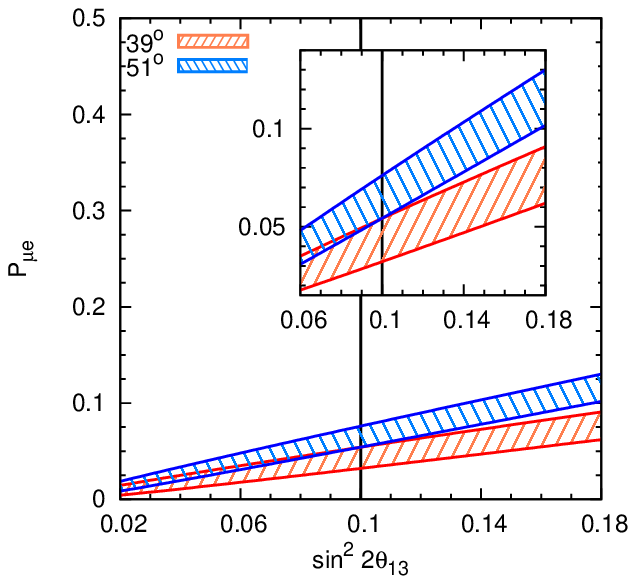, width=7.6cm,height=6.875cm, bbllx=80, bblly=50, bburx=265, bbury=226,clip=}
\hspace*{-0.2in}
\epsfig{file=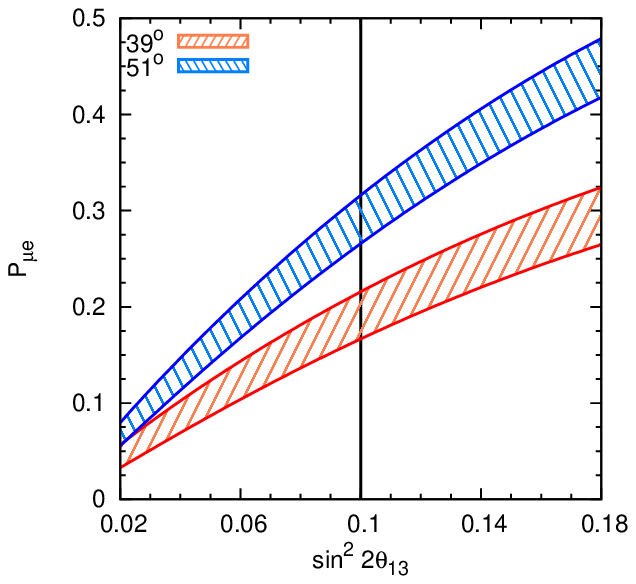, width=7.6cm,height=6.875cm, bbllx=80, bblly=50, bburx=265, bbury=226,clip=} \\
\epsfig{file=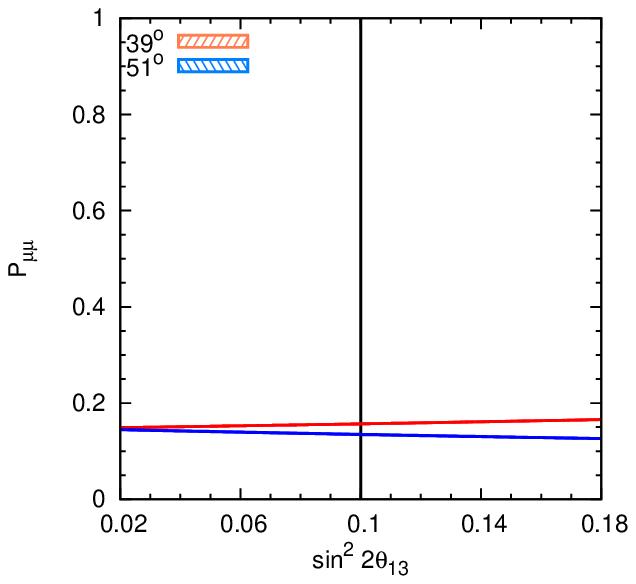, width=7.6cm,height=6.875cm, bbllx=80, bblly=50, bburx=265, bbury=226,clip=}
\hspace*{-0.2in}
\epsfig{file=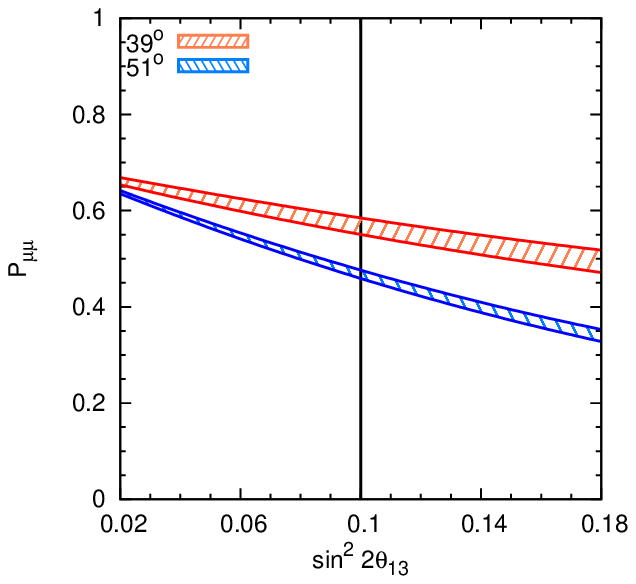, width=7.6cm,height=6.875cm, bbllx=80, bblly=50, bburx=265, bbury=226,clip=} \\
\vspace{-0.2in}
\caption{\small Behaviour of the muon survival and oscillation probabilities 
as a function of 
$\sin^2 2\theta_{13}$ showing the 
$\theta_{23}$ octant degeneracy and its breaking. The left panels 
are for the \nova\ peak energy and baseline (2 GeV, 812 Km), while the 
right panels are for a typical atmospheric neutrino energy and baseline (6 GeV, 5000 Km). 
The top row denotes the appearance channels $P_{\mu e}$ and $P_{e\mu}$, 
while the bottom row denotes the disappearance channel $P_{\mu\mu}$.
The values of oscillation parameters chosen are 
$\theta_{23}^{tr} = 39^o$, $\theta_{23}^{wrong} = 51^o$. The bands denote a variation 
over the full range of the phase $\delta_{CP}$. The inset shows the region of 
separation of the bands near $\sin^2 2\theta_{13} = 0.1$. }
\label{Pvsth13_delcptest}}

%In the fixed parameter case, i.e. if we assume that $\theta_{13}$ and 
%$\sin^2 2\theta_{23}$ are known exactly, it is easy to check the octant 
%sensitivity at the probability level itself. One must of course, allow for 
%variation in $\delta_{CP}$ in its full range, since its value is unknown. 
Figure~\ref{Pvsth13_delcptest} depicts the probabilities $\pmue$ and $\pmumu$ as a 
function of $\sin^2 2\theta_{13}$  for 
$\theta_{23}^{tr} = 39^o$ and $\theta_{23}^{wrong} = 51^o$.
% with a fixed baseline L$=$812 Km 
%and E$=$2 GeV corresponding to the baseline and peak energy of the 
%NOvA experiment (left panels) and with a sample atmospheric neutrino baseline 5000 Km and energy 6 GeV (right panels). 
The bands show the probability range in each case 
when $\delta_{CP}$ is varied over its full range (0 to 2$\pi$). 
%In the case of NOvA and T2K-like baselines, matter effects arise only 
%at subleading order as discussed above, and hence the manifestation 
%of the octant degeneracy with $\theta_{13}$ and $\delta_{CP}$ is more 
%prominent. 
For a given fixed value of $\sin^2 2\theta_{13}$, the distinction between 
$\theta_{23}$ in the two octants can be gauged from the separation of 
the two bands along the relevant vertical line. 
The left panels show that for $\pmue$, the $\delta_{CP}$ 
bands overlap and the two hierarchies cannot be distinguished till 
nearly $\sin^2 2\theta_{13} = 0.1$. 
The inset in the upper left panel shows the region of separation of the 
bands near $\sin^2 2\theta_{13} = 0.1$ in detail. 
Hence the knowledge of the 
parameter $\theta_{13}$ upto its current level of precision becomes 
crucial, since a $\theta_{13}$ range including lower values would 
wash out the octant sensitivity
derivable from such experiments due to the combined degeneracy 
with $\theta_{13}$ and $\delta_{CP}$. 
For $\pmumu$, 
the intrinsic degeneracy predominates
%and the separation between the octants is statistically small, 
and the effect
of $\delta_{CP}$ variation is insignificant. 

For the 5000 km baseline, due to the resonant matter effects breaking 
the octant degeneracy at the leading order, both $\pmue$ and $\pmumu$ 
show a wider separation between the opposite-octant bands, even for
small values of $\theta_{13}$. This is due to the 
$\sin^2 \theta_{23} \sin^2 2 \theta_{13}^{\rm{m}}$ ($\sin^4 \theta_{23} 
\sin^2 2 \theta_{13}^{\rm{m}}$) term in $\pmue$ ($\pmumu$). 
The $\delta_{CP}$ bands in the right-hand panels are much wider 
because of the enhancement of the subleading terms due to matter effects. 
However, the enhancement is more for the  leading order 
term 
which alleviates the  
degeneracy with $\delta_{CP}$.
% demonstrating clear signals of 
%octant sensitivity. 

%\end{itemize}

\FIGURE{ 
\epsfig{file=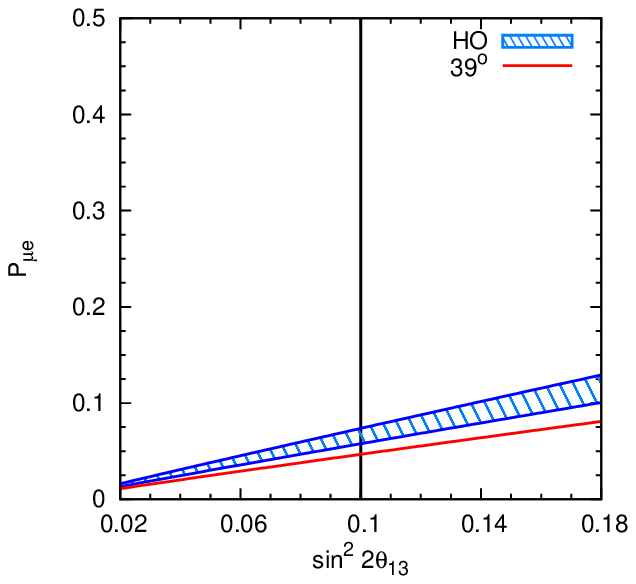, width=7.6cm,height=6.875cm, bbllx=80, bblly=50, bburx=265, bbury=226,clip=}
\hspace*{-0.2in}
\epsfig{file=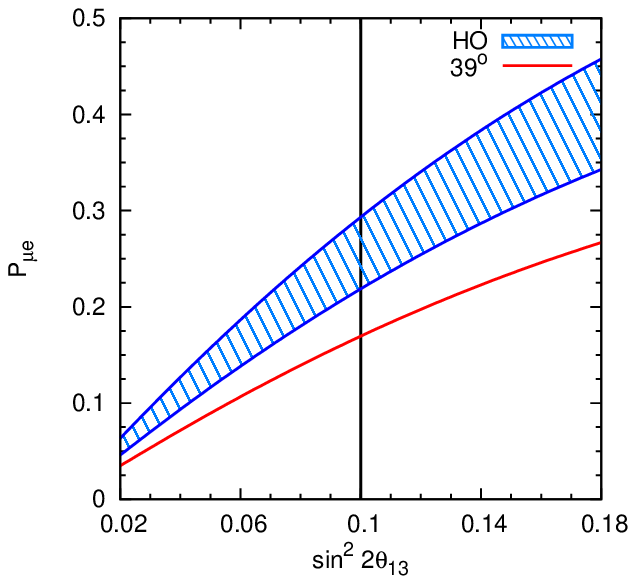, width=7.6cm,height=6.875cm, bbllx=80, bblly=50, bburx=265, bbury=226,clip=} \\
\epsfig{file=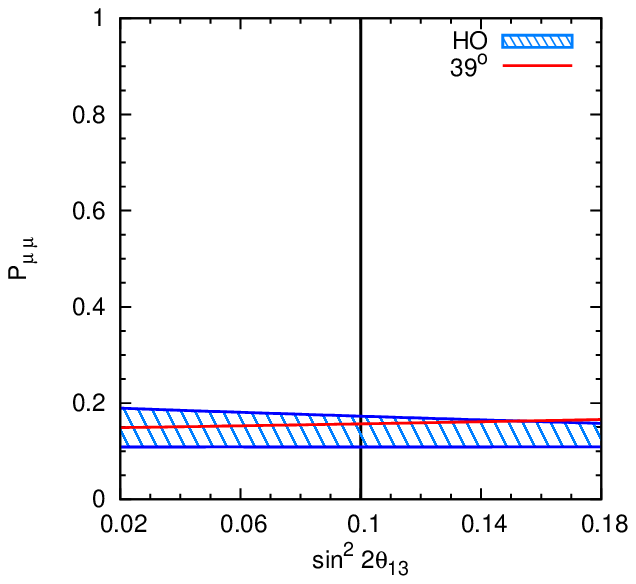, width=7.6cm,height=6.875cm, bbllx=80, bblly=50, bburx=265, bbury=226,clip=}
\hspace*{-0.2in}
\epsfig{file=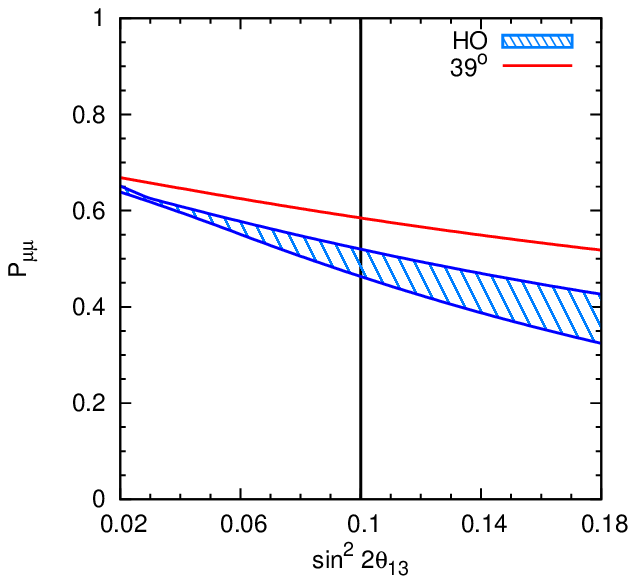, width=7.6cm,height=6.875cm, bbllx=80, bblly=50, bburx=265, bbury=226,clip=} \\
\vspace{-0.2in}
\caption{\small Same as Figure~\ref{Pvsth13_delcptest} with a fixed true and 
test $\delta_{CP}=0$ 
and true $\theta_{23}^{tr} = 39^o$, with the band denoting a variation over 
the full allowed range of
$\theta_{23}^{wrong} = 45^o$ to $54^o$ in the wrong octant.}
\label{Pvsth13_th23test}
}

\FIGURE{ 
\epsfig{file=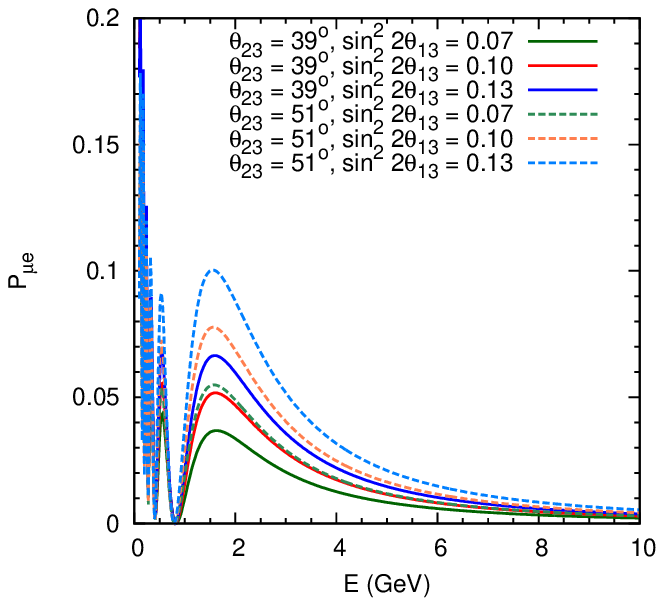, width=7.4cm,height=6.875cm, bbllx=80, bblly=50, bburx=265, bbury=226,clip=}
\epsfig{file=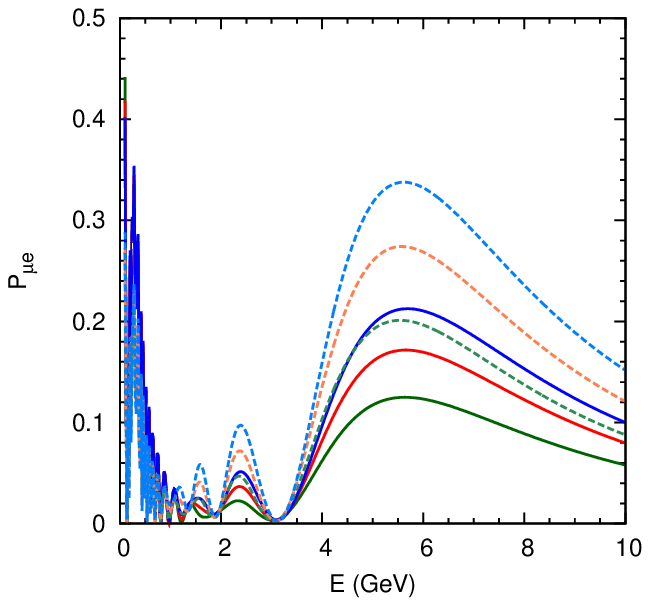, width=7.4cm,height=6.875cm, bbllx=80, bblly=50, bburx=265, bbury=226,clip=} \\
\epsfig{file=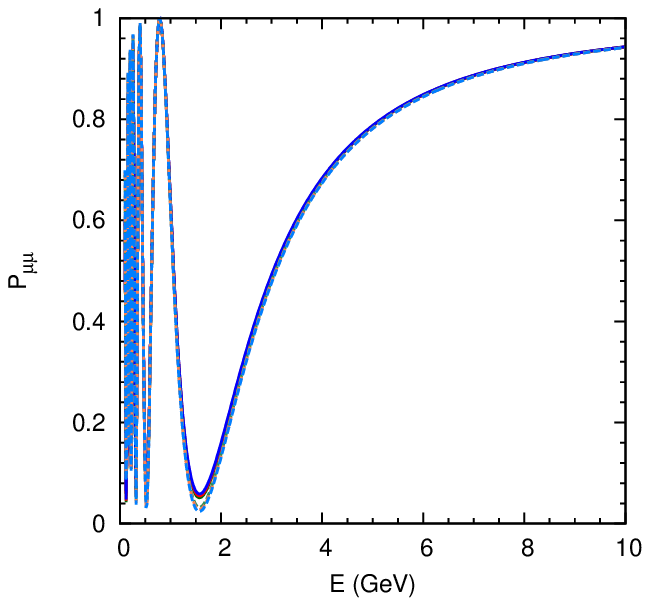, width=7.4cm,height=6.875cm, bbllx=80, bblly=50, bburx=265, bbury=226,clip=}
\epsfig{file=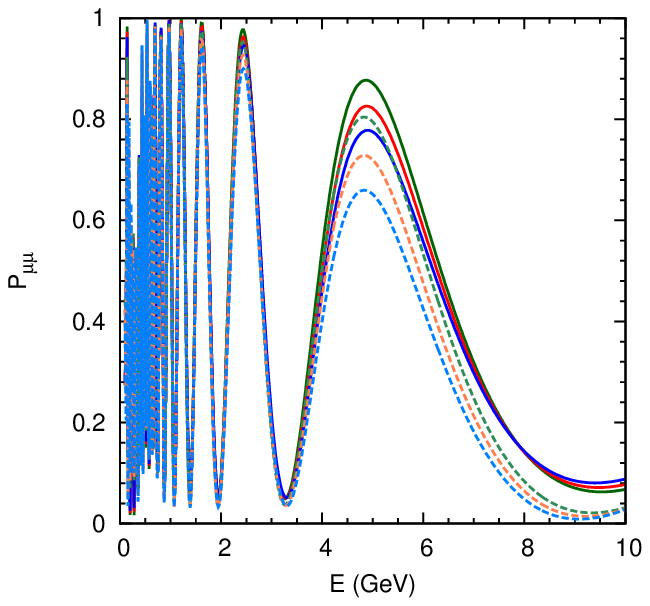, width=7.4cm,height=6.875cm, bbllx=80, bblly=50, bburx=265, bbury=226,clip=} \\
\vspace{-0.1in}
\caption{\small Energy spectra of the probabilities $\pmue$ and $\pmumu$ for the 
\nova\ baseline (left panels) and for a sample atmospheric neutrino baseline 5000 Km 
(right panels). The figure shows the variation in the probabilities in each case 
when $\sin^2 2\theta_{13}$ is varied over three values in the current allowed range, 
fixing the true and test $\theta_{23}$ values $\theta_{23}^{tr} = 39^o$ and 
$\theta_{23}^{wrong} = 51^o$. $\delta_{CP}$ is fixed to 0.}  
\label{PvsE_th13}
}

\FIGURE{ 
\epsfig{file=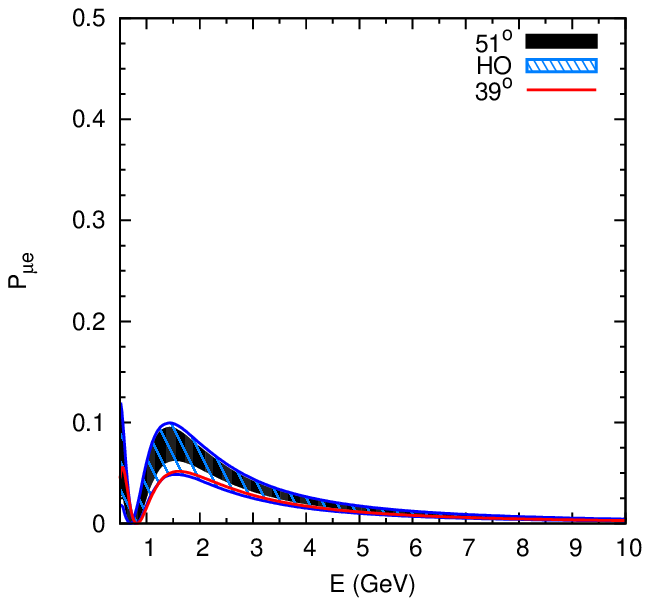, width=7.4cm,height=6.875cm, bbllx=80, bblly=50, bburx=265, bbury=226,clip=}
\epsfig{file=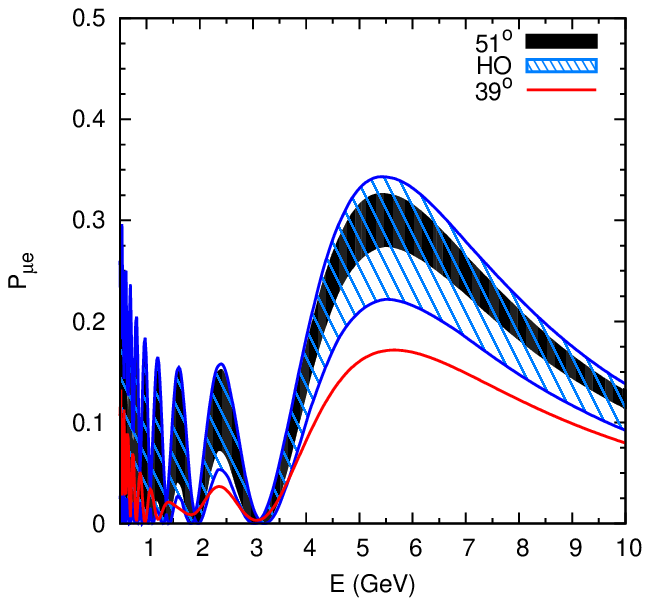, width=7.4cm,height=6.875cm, bbllx=80, bblly=50, bburx=265, bbury=226,clip=} \\
\epsfig{file=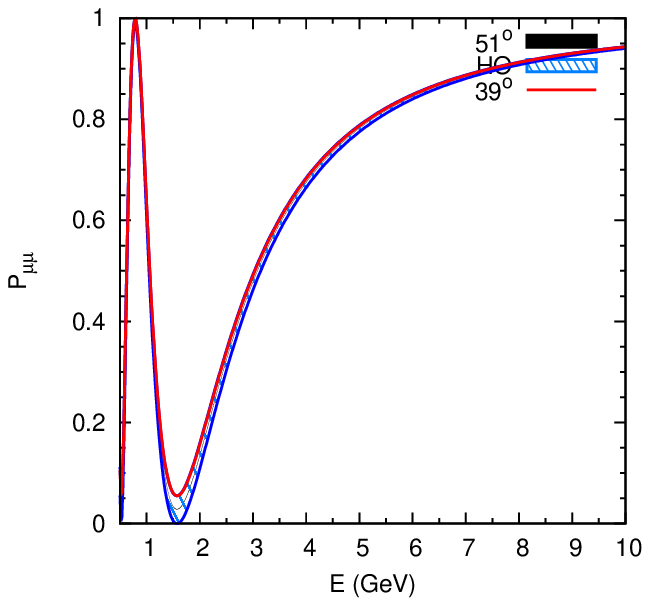, width=7.4cm,height=6.875cm, bbllx=80, bblly=50, bburx=265, bbury=226,clip=}
\epsfig{file=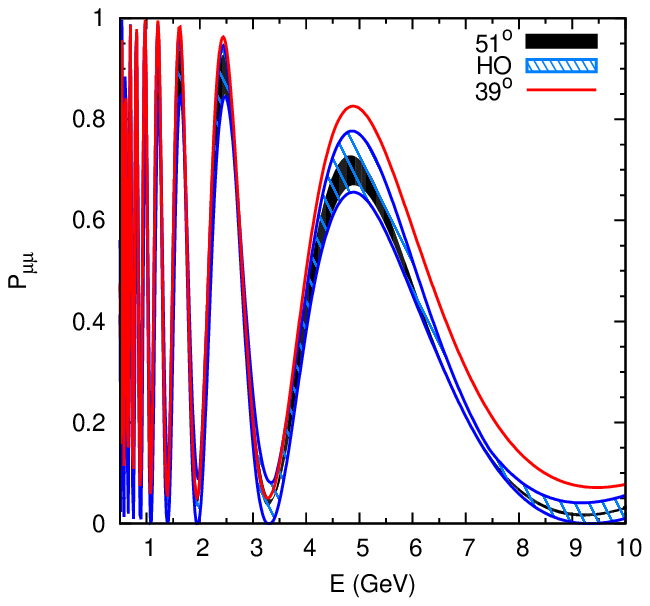, width=7.4cm,height=6.875cm, bbllx=80, bblly=50, bburx=265, bbury=226,clip=} \\
\vspace{-0.1in}
\caption{\small Energy spectra of the probabilities $\pmue$ and $\pmumu$ for the 
\nova\ baseline (left panels) and for a sample atmospheric neutrino baseline 5000 Km 
(right panels). The figure shows the variation in the probabilities in each case 
when $\theta_{23}^{wrong}$ is varied over the entire
allowed range in the wrong octant, as well as varying the test $\delta_{CP}$, 
fixing the true values $\theta_{23}^{tr} = 39^o$ and $\delta_{CP}^{tr} = 0$ 
in the solid curve.  The spread due to variation of both test 
parameters is denoted by the blue band. The black band shows
the variation with test $\delta_{CP}$ for a fixed $\theta_{23}^{wrong} = 51^o$.} 
\label{PvsE_th23delcptest}
}

%\begin{figure*}[t]
%\vskip 1cm
%\includegraphics[scale=.55]{}
%\includegraphics[scale=.6]{prob1}
%\includegraphics[scale=1.1]{oct_degenn_dm31_dcp0.eps}
%\hspace*{-0.8in}
%\includegraphics[scale=1.1]{oct_degenn_dm31_dcp90.eps}
%%\includegraphics[scale=0.7]{octsens}
%\caption{\small Behaviour of the oscillation probability $P_{\mu e}$ 
%for $\theta_{23}$ values in the two octants as a function of energy 
%for the NOvA baseline 812 Km, with a fixed $\sin^2 2\theta_{13} = 0.1$ and  
%$\dcp = 0^o$(left panel) or $\dcp = 90^o$(right panel). The values chosen 
%are $\theta_{23}^{tr} = 39^o$, 
%$\theta_{23}^{wrong} = 51^o$. The bands denote a variation over 
%the $3\sigma$ allowed range of $\Delta m_{31}^2$.
% }
%\label{pmuevsE_812}
%\end{figure*}

Figure~\ref{Pvsth13_th23test} shows the probabilities $\pmue$ and $\pmumu$ as a 
function of $\sin^2 2\theta_{13}$. We have held fixed, true and test $\delta_{CP}=0$ 
and true $\theta_{23}^{tr} = 39^o$, with the band denoting a variation over the 
full allowed range of
$\theta_{23}^{wrong} = 45^o$ to $54^o$ in the wrong octant. 
Thus this figure reveals the effect of the uncertainty in the 
measurement of $\theta_{23}$ in the determination of octant, for a given value 
of $\sin^2 2\theta_{13}$. For the \nova\ baseline, the survival probability 
$\pmumu$ shows an overlap of the
test $\theta_{23}$ band with the true curve, while in the probability $\pmue$ 
there is a small separation. Figures~\ref{Pvsth13_delcptest} and 
\ref{Pvsth13_th23test} thus indicate that for the \nova\ baseline, the 
octant sensitivity from $\pmue$ is more affected
by the uncertainty in $\delta_{CP}$ and less by the test $\theta_{23}$ 
variation, while for $\pmumu$ the opposite is true. 
%{\bf{Thus combining the appearance and disappearance channels
%in experiments like \nova and T2K can enhance octant sensitivity due
%to the tension in the behaviour of the two probabilities as a function of $\delta_{CP}$ and $\theta_{23}$}}.  
%Further, the effect of uncertainty in 
%$\delta_{CP}$ can be reduced by including data from a short baseline experiment 
%such as T2K. These claims will be substantiated later, when the combined octant 
%sensitivity of NOvA and T2K will be discussed}. 
The plots for the atmospheric neutrino baseline show a clear breaking of the 
octant degeneracy 
in both $\pmue$ and $\pmumu$ even for small values of $\theta_{13}$, 
indicating that the octant sensitivity from the atmospheric neutrino
signal is stable against the variation of both $\delta_{CP}$ and the test 
value of $\theta_{23}$, even for small values of $\theta_{13}$. 

Having discussed octant sensitivity in the context of fixed $\theta_{13}$, 
we now come to the question of uncertainty in this parameter. 
The current reactor measurements of $\sin^2 2\theta_{13}$ 
have measured this parameter with a precision of $0.01$  and
hence the effect of $\theta_{13}$ uncertainty on octant degeneracy 
is largely reduced. 
In Figure~\ref{PvsE_th13}, the energy spectra of the probabilities $\pmue$ and 
$\pmumu$ are plotted. 
The figure shows the variation in the probabilities 
in each case when $\sin^2 2\theta_{13}$ is varied over three values 
-- $\sin^2 2\theta_{13} = 0.07, 0.1$ and $0.13$ 
covering the current allowed range, and for  two illustrative 
values of $\theta_{23}$ --  $39^o$ and $51^o$ in opposite octants. 
The value  of $\delta_{CP}$ 
is  fixed to be $0$. The figure indicates that the 
separation of the LO and HO  curves and hence the octant sensitivity 
depends on the value of $\sin^2 2\theta_{13}$ in opposite ways 
depending on whether the true $\theta_{23}$ value lies in the higher or lower 
octant. In $\pmue$, for  $\theta_{23}^{tr}=39^o$, lower values 
of true $\sin^2 \theta_{13}$ can give a higher octant sensitivity
since they are 
more separated from the band of variation of the 
probability over the whole range of 
test
$\sin^2 2\theta_{13}$, and as 
true $\sin^2 2\theta_{13}$ increases the degeneracy with the 
wrong $\theta_{23}$ band becomes more prominent. For $\theta_{23}^{tr} = 51^o$, 
the opposite is true, 
i.e. higher values of $\sin^2 2\theta_{13}^{tr}$ have a better separation with 
the wrong $\theta_{23}$ band. For example, for the \nova\ baseline, 
if $\theta_{23}^{tr} = 39^o$, the $\pmue$ curve for $\sin^2 2\theta_{13}^{tr} = 0.07$ 
is well-separated from the band for $\theta_{23}^{wrong} = 51$, 
while the curve for $\sin^2 2\theta_{13}^{tr} = 0.1$ is seen to be just separated 
from it, and the $\sin^2 2\theta_{13}^{tr} = 0.13$ curve lies entirely 
within the band of test $\sin^2 2\theta_{13}$ variation with the wrong octant. But 
if $\theta_{23}^{tr} = 51^o$, the $\sin^2 2\theta_{13}^{tr} = 0.07$ curve suffers 
from the degeneracy while the curves for $\sin^2 2\theta_{13}^{tr} = 0.1$ and 
upwards lie clearly outside the $\theta_{23}^{wrong}$ band.    
For $\pmumu$, the effect of $\theta_{13}$ on the separation between the 
opposite octant bands
is less, since the behaviour is governed by the intrinsic octant degeneracy.

For the 5000 km baseline, due to strong matter effects, the separation of 
the true and wrong $\theta_{23}$ bands is much better, and only the highest
(lowest) 
values of $\sin^2 2\theta_{13}^{tr}$ suffer from a degeneracy with the wrong 
octant 
for $\pmue$ if the true octant is lower (higher). For $\pmumu$, the behaviour is 
reversed.

Figure~\ref{PvsE_th23delcptest} again depicts the energy spectra of the probabilities 
$\pmue$ and $\pmumu$.
This figure shows the variation in the probabilities in each case when 
$\theta_{23}^{wrong}$ is varied over the entire
allowed range in the wrong octant, as well as varying the test $\delta_{CP}$, 
fixing $\theta_{23}^{tr} = 39^o$ and $\delta_{CP}^{tr} = 0$. The solid black band denotes
the variation with test $\delta_{CP}$ for a fixed $\theta_{23}^{wrong} = 51^o$.

For the \nova\ baseline, the test probability bands show an almost complete overlap 
with the $\theta_{23}^{tr} = 39^o, \delta_{CP}^{tr} = 0$ curve. However, 
the octant sensitivity may still be retained due to spectral 
information.
%since no two sets of true and test parameters
%suffer from an exact coincidence of the corresponding probability spectra. 
So there always exist specific energy regions 
and bins from which the octant sensitivity can be derived.     

At $5000$ km, the minimum separation between octants does not 
occur at or near $\theta_{23}^{wrong} = 90^o - \theta_{23}^{tr} = 51^o$
as can be seen from the solid black shaded region. 
The edge of the striped blue band corresponding to 
some other value of $\theta_{23}$ is closest to the `true' curve 
and well-separated from it. This shows that there is octant sensitivity 
for this baseline even after including the uncertainty in $\delta_{CP}$ and  
$\theta_{23}$. 
The $\theta_{23}^{wrong}$ for which 
the minimum separation is likely to occur  will be further discussed  
in the context of Figure~\ref{octant_testth23}.
%In Figure \ref{pmuevsE_812}, the energy spectra of the muon oscillation and 
%survival probabilities are plotted for fixed $\sin^2 2\theta_{13} = 0.1$, 
%for a baseline of 812 Km, with bands denoting a variation over the 
%3$\sigma$ range of the atmospheric mass-squared difference 
%$\Delta m_{31}^2$. 
%and any degeneracy 
%with this parameter occurs well below the peak energy range of 
%the experiments}}. 

%\begin{figure} 
%\includegraphics[width =11.8cm,height=9.2cm,clip=,angle=0]{probarray_oct7000_new.eps}
%\caption{\small Dependence of the neutrino survival and 
%oscillation probabilities $P_{\mu\mu}$ and $P_{\mu e}$ on the 
%octant of $\theta_{23}$, for a baseline of 7000 km, for $\sin^2 2\theta_{13} = 0.1$, 
%$\dcp = 0^o$.}
%\label{prob7000}
%\end{figure} 

\section{Analysis and Results} 
\label{sec:analysis}

In this section we present the results of our 
analysis for T2K, \nova\ and atmospheric 
experiments. We also give results for octant sensitivity
when the results from both type of experiments are combined.  

%{\bf{We fix the true values of the parameters 
% respectively}}.
%{\bf true values of th23, del31 ? do we take delatm and thatm true value ?}
%The best measurement of the atmospheric parameters till date is from MINOS 
%and SK. Recently, different global fits have favoured the lower octant 
%or the hight octant at the $1-2\sigma$ level. \nova and T2K are 
%expected to improve the precision on these parameters. However, in this work, 
%we will consider the effect of adding information from \nova and T2K on 
%the octant sensitivity of atmospheric neutrino experiments. Therefore, we do not use 
%the projected ranges of the atmospheric parameters coming from \nova and T2K. We only 
%take into acccount the rather conservative parameter range coming from prior 
%measurement of atmospheric parameters from MINOS \cite{minos_2012}. 
In Refs.~\cite{dm31_defn,th23_defn}, it has been shown that the 
atmospheric parameters $\dmatm$ and $\tmm$ measured in MINOS are related to the 
oscillation parameters in nature, $\Delta m_{31}^2$ and $\theta_{23}$ using the following 
non-trivial transformations:
\begin{equation}
 \sin\theta_{23} = \frac{\sin\tmm}{\cos\theta_{13}} \ ~,
\end{equation}
\begin{equation}
 \Delta m_{31}^2 = \dmatm + (\cos^2\theta_{12} - \cos\delta\sin\theta_{13}\sin2\theta_{12}
\tan\theta_{23})\Delta m_{21}^2\ ~.
\end{equation}
These transformations become significant in light of the moderately large measured value 
of $\theta_{13}$. Therefore, 
in order to avoid getting an erroneous estimate of octant sensitivity, we take 
these `corrected' definitions into account. 
Thus, in calculating oscillation probabilities, 
we use the corrected parameters $\Delta m_{31}^2$ and $\theta_{23}$, after allocating
the measured values to $\dmatm$ and $\tmm$.

Our analysis procedure consists of simulating the 
experimental data for some specific values of 
the known oscillation parameters known as the `true values'. 
The experimental data is generated for a fixed hierarchy
and for the following fixed values of the parameters:
\begin{eqnarray}
(\Delta m^2_{21})^{tr} & = & 7.6 \times 10^{-5} \textrm{ eV}^2 \nonumber \\
(\sin^2\theta_{12})^{tr} & = & 0.31 \nonumber \\
(\sin^2 2\theta_{13})^{tr} & = & 0.1 \\
(\dmatm)^{tr} & = & 2.4 \times 10^{-3} \textrm{ eV}^2 \nonumber ~,
\label{eq:truevals}
\end{eqnarray}
and specific values
of $\tmm^{tr}$ and $\delta_{CP}^{tr}$. 
In the theoretical predictions which are fitted to the 
simulated experimental data, the `test' parameters are marginalized 
over the following ranges:
\begin{eqnarray}
\delta_{CP} & \in & [0,2\pi)  \nonumber \\
\theta_{\mu \mu} & \in & \left\{
\begin{array}{ll}
(35^o,45^o) & \textrm{(true higher octant)} \\ 
(45^o,55^o) & \textrm{(true lower octant)} 
\end{array}
\right.
\nonumber \\
\sin^2 2\theta_{13} & \in & (0.07,0.13) ~. \nonumber \\
\label{eq:rangevals}
\end{eqnarray}
$\Delta m^2_{21}$ and $\sin^2\theta_{12}$ are fixed to their true values
since their effect is negligible. Also, after verifying that
the effect of a marginalization over $\Delta m_{31}^2$ 
is minimal, we have fixed $\dmatm$ to its true value
for computational convenience. 

In our calculation, the hierarchy is assumed to be known in all the atmospheric neutrino 
experiments, since the time-scale involved would ensure that the hierarchy
is determined before any significant octant sensitivity is achievable. 
For \nova/T2K, results are given both with and without prior 
knowledge of hierarchy, i.e. marginalizing over the test hierarchy. 

Priors are taken in terms of the measured quantities $\sin^2 2\theta_{13}$ and 
$\sin^2 2\tmm$ as follows:

%\begin{eqnarray}
%{\rm{\chi^2_{prior}}}  & = & 
%\left(\frac{|\da|^{\mathrm{true}} - |\da|}
%{\sigma(|\da|)}\right)^2  \right. \nonumber \\
%& & \left. + \left(\frac{{\sin^2 2\theta_{23}^{\mathrm{true}}} -
%\sin^2 2\tmm^{\rm{true}}}
%{\sigma(\sin^2 2\tmm)}\right)^2
%+ \left(\frac{{\sin^2 2\theta_{13}^{\mathrm{true}}} - \sin^2 2\theta_{13}}
%{\sigma(\sin^2 2\theta_{13})}\right)^2 \right]
%\label{chisqprior}
%\end{eqnarray}
%}}

\begin{equation}
\chi^2_{prior} = 
\left(\frac{{\sin^2 2\tmm^{\mathrm{true}}} -
\sin^2 2\tmm}
{\sigma(\sin^2 2\tmm)}\right)^2
+ \left(\frac{{\sin^2 2\theta_{13}^{\mathrm{true}}} - \sin^2 2\theta_{13}}
{\sigma(\sin^2 2\theta_{13})}\right)^2
\end{equation}
with the 1$\sigma$ error ranges as 
$\sigma_{\sin^2 2\theta_{\mu\mu}} = 5\%$ and
$\sigma_{\sin^2 2\theta_{13}} = 0.01$ unless otherwise stated. The latter is the error
on $\theta_{13}$ quoted recently by Double Chooz, Daya Bay and RENO 
\cite{DC,daya,reno}. 

%\section{Experimental specifications}
%\subsection{Atmospheric neutrino detectors} 
%{{
%For our study of octant sensitivity in atmospheric neutrino experiments, we look at the following set-ups:

\subsection{\nova\ and T2K}

\FIGURE{
\includegraphics[width =\textwidth,clip=,angle=0]{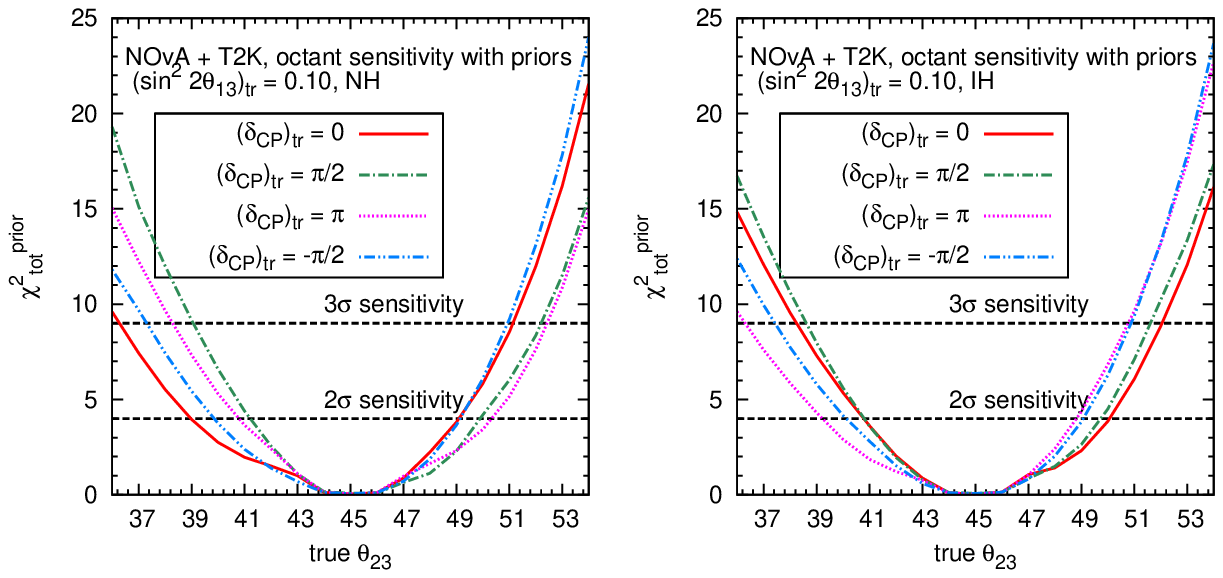}
\vspace{-0.4in}
\caption{\small Marginalized octant sensitivity from a combination of \nova\ and T2K,
for the case of normal and inverted mass hierarchy. The test hierarchy has been fixed to be
the same as the true hierarchy in this case. In this figure
priors have been added.}
\label{octant_NovaT2K}
}

\FIGURE{
\includegraphics[width =\textwidth,clip=,angle=0]{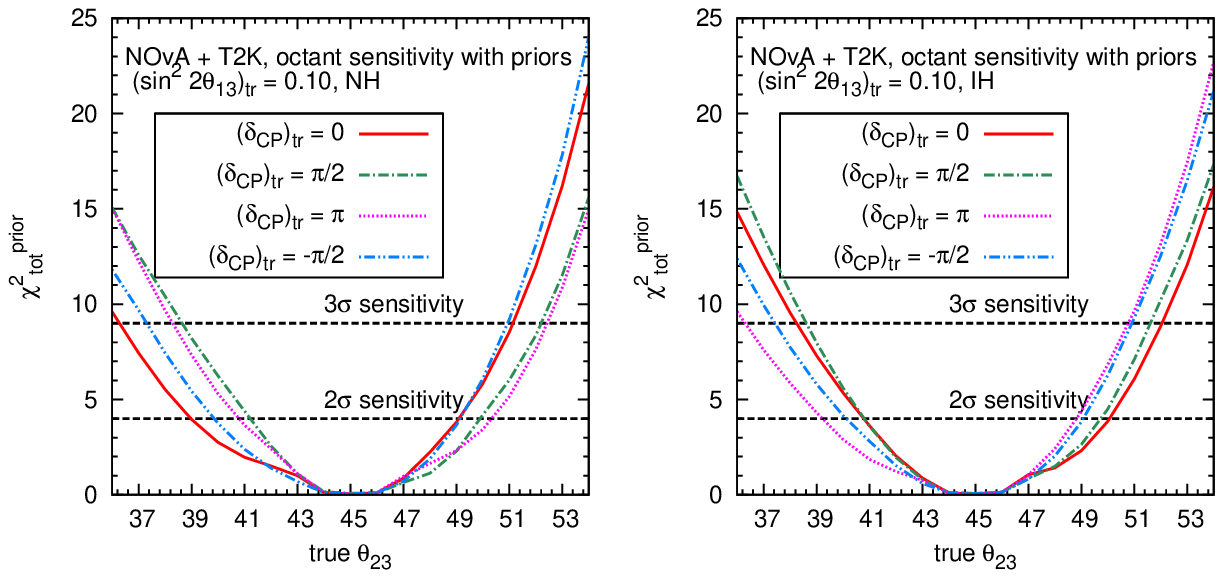}
\vspace{-0.4in}
\caption{\small Marginalized octant sensitivity from a combination of \nova\ and T2K,
for the case of normal and inverted mass hierarchy. The test hierarchy has been left free
in this case. In this figure priors have been added.}
\label{octant_NovaT2K_marghier}
}

We simulate the current generation of long baseline experiments
\nova\ and T2K, using 
the GLoBES package~\cite{globes,globes2} and its associated data 
files~\cite{globes_xsec,globes_xsec2}. 
For T2K, we assume a 3 year run with neutrinos alone, running with beam power of 
$0.77$ MW throughout. (We choose this low running time for T2K to compensate for the 
fact that their beam power will be increased to its proposed value over a 
period of a few years~\cite{t2k_beampower}.)
The energy resolutions and backgrounds are taken from 
Refs.~\cite{t2k_globes,t2k_globes2,t2k_globes3,t2k_globes4,t2k_globes5}. 
The detector mass is taken to be $22.5$ kT. For \nova, we consider the set-up as 
described in Refs.~\cite{nova_reopt,nova_reopt2}, which is re-optimized 
for the moderately 
large measured value of $\theta_{13}$. The $14$ kT detector receives 
a neutrino and 
antineutrino beam for 3 years each, from the NuMI beam. 

Figures~\ref{octant_NovaT2K} and \ref{octant_NovaT2K_marghier} depict the octant 
sensitivity from a combination of \nova\ and 
T2K marginalized over the test parameters and with added priors.
The $\chi^2$ is defined as 
\bea
{\rm{(\chi^2_{tot})^{prior}}} = {\rm{min(\chi^2_{NOvA} + \chi^2_{T2K} + \chi^2_{prior}) }}
\eea
Here (and elsewhere), `min' denotes a marginalization over the test parameters as outlined 
above. In Figure~\ref{octant_NovaT2K}, the neutrino mass hierarchy is assumed to be 
known in each case, and in Figure~\ref{octant_NovaT2K_marghier}
the mass hierarchy is taken to be unknown and therefore marginalized over. 
These plots are done 
for four specific true values of the CP phase, $\dcp = 0, \pi/2, \pi$ and -$\pi/2$. 
%like Figures \ref{octant_LAr} and \ref{octant_ICAL}.
%As will be shown at the end of this section, the effect of varying $\dcp$ is more
%prominent in the case of NOvA/T2K compared to atmospheric neutrino experiments. 

\begin{figure*}[t]
\includegraphics[width=\textwidth]{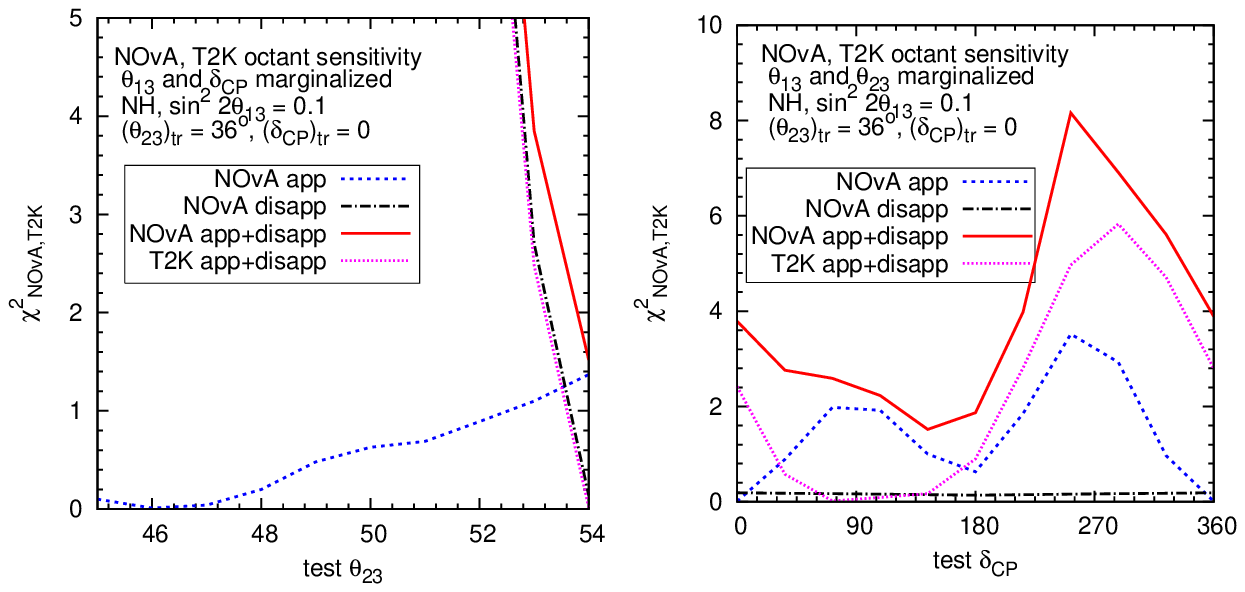}
\vspace{-0.3in}
\caption{{\small{Octant sensitivity from \nova\ appearance and disappearance 
channels and their combination
and from T2K as a function of test $\theta_{23}$ (left panel) and test 
$\delta_{CP}$ (right panel).  
$\theta_{13}$ and $\delta_{CP}$ are marginalized in the left panel
and $\theta_{13}$ and $\theta_{23}$ are marginalized in the right panel.
}}}
\label{NovaT2K_testth23}
\end{figure*}

Table~\ref{tablelimits_NovaT2K} lists the values of the octant sensitivity from \nova\ 
and T2K individually and from the combination for a specific value $\delta_{CP}^{tr} = 0$, 
and a set of true $\theta_{23}$ values from both octants, without and with added priors, 
to depict the relative contributions from the two experiments.
Figure~\ref{NovaT2K_testth23} shows how the octant 
$\chi^2$ for $\theta_{23}^{tr} = 36^o$ varies with the test values 
of $\theta_{23}$ and $\delta_{CP}$ for the appearance and disappearance channels of 
\nova\ and for T2K. In the figure, $\theta_{13}$ and $\delta_{CP}$ are marginalized 
in the left panel and $\theta_{13}$ and $\theta_{23}$ are marginalized in the right 
panel. Normal hierarchy is assumed in the figure.
The following features are observed in the sensitivity behaviour
of the two experiments and the appearance/disappearance channels in each case:

\begin{table}
%\TABLE[!h]{\centerline{
\begin{center}
%\TABULAR
\begin{tabular}{|c || c | c | c | c | c |} \hline
        {{$\theta_{23}^{tr}$}} & {{$\chi^2$ (\nova)}}
& {{$\chi^2$ (T2K)}} &  {{$\chi^2$ (\nova}} &  
{{$\chi^2$ (\nova}} &  {{$\chi^2$ (\nova}}
        \\
   & & & {{$+$T2K)}} & {{$+$T2K$+$prior)}} & {{$+$T2K$+{\rm{prior_{n}}}$)}}
        \\
        \hline
    36 & 1.5 (4.1) & 0.0 (0.8) & 1.7 (5.8) & 9.6 (14.8) & 17.5 (26.7) \\ \hline
     39 & 0.2 (0.4) & 0.0 (0.1) & 0.3 (0.6) & 3.9 (7.3) & 6.3 (12.0) \\ \hline
     41 & 0.1 (0.1) & 0.0 (0.0) & 0.1 (0.1) & 1.9 (3.6) & 2.4 (5.4) \\ \hline
        43  &  0.1 (0.1) & 0.0 (0.0) & 0.1 (0.1) & 1.0 (0.8) & 1.3 (1.1) \\ \hline
         47 & 0.0 (0.1) & 0.0 (0.0) & 0.1 (0.1) & 0.8  (1.0) & 1.0 (1.3) \\ \hline
  49 & 0.2 (0.3) & 0.0 (0.0) & 0.3 (0.5) & 3.8 (2.3) & 5.4 (3.1) \\ \hline
  51 & 2.3 (1.1) & 0.2 (0.1) & 2.9 (1.5) & 8.5  (6.0) & 13.0 (8.2) \\ \hline
  54 & 11.0 (6.8) & 2.8 (0.5) & 14.7 (9.4) & 21.5 (16.1) & 32.4 (22.6) \\ \hline
%     44  & 0.0 &  0.0 & 0.1   \\ \hline
%        0.12  & 15.5 &  8.3 & 12.1
%\\ \hline
%        0.15  & 6.1 & 7.1 & 11.8 & 16.9

%       $\sin^2 \theta_{23} $ & 0.50 &  0.34 -- 0.68 \\ \hline
%       $\sin^2 \theta_{13} $ & 0.00 &  $\le$ 0.047
           \hline
\end{tabular}
\caption{\small Marginalized octant sensitivity from \nova, T2K and a
combination of
the two experiments for $\delta_{CP}^{tr} = 0$ and $\sin^2 2\theta_{13}^{tr} = 0.1$
without added priors (first three columns) and with added priors
(fourth and fifth columns). 'prior' denotes the present prior of $\sigma_{\sin^2 2\theta_{13}} = 0.01$ and '${\rm{prior_n}}$' denotes a projected prior of  $\sigma_{\sin^2 2\theta_{13}} = 0.005$. Here normal hierarchy (inverted hierarchy) is
assumed.}
\label{tablelimits_NovaT2K}
\end{center}
\end{table}

\begin{enumerate}

\item Figure~\ref{NovaT2K_testth23} (left panel) shows that the $\chi^2$ 
minima for the \nova\ disappearance channel occur near
$\theta_{23}^{test} = \pi/2 - \theta_{23}^{tr}$, because of the predominant
dependence on $\sin^2 2\theta_{23}$, while for the appearance channel the minima
occur near $\theta_{23}^{test} = 45^o$, because of the $\sin^2 \theta_{23}$ dependence.
In the combination of \nova\ appearance and disappearance channels
(red solid curve in the figure), 
the $\chi^2$ minima are near $\theta_{23}^{test} = \pi/2 - \theta_{23}^{tr}$,
following
the behaviour of the disappearance channel $\chi^2$, but the values are
enhanced
due to the contribution from the appearance $\chi^2$ values at that point.

\item 
The $\chi^2$ values are asymmetric across the lower and higher octants,
as can be seen in Table ~\ref{tablelimits_NovaT2K}.
For example, $\chi^2_{min}$ is about 1.5 for 
\nova\ (appearance $+$ disappearance) at
$\theta_{23}^{tr} = 36^o$ with $\sin^2 2\theta_{13}^{tr} = 0.1, \delta_{CP}^{tr} = 0$.
For $\theta_{23}^{tr} = 54^o$, $\chi^2_{min}$ goes
up to about 11 for \nova\ (app $+$ disapp) for  the same value of $\delta_{CP}^{tr}$.

\item  The $\chi^2$ values  for \nova\ and T2K strongly depends on the 
true value of $\delta_{CP}$ . We find 
for $\delta_{CP}^{tr} = \pi/2$,
$\chi^2_{min}$ for \nova\ (app $+$ disapp) is 5.6 for $\theta_{23}^{tr} = 36^o$ 
and 7.5 for $\theta_{23}^{tr} = 54^o$.
This point will be illustrated in more detail later in the context of 
Figures~\ref{octant_dcpt23} and \ref{octant_contdcpt23}. 

%This leads to a strong tension between the $\chi^2$ behaviour as a
%function of $\theta_{23}^{test}$, so that the combination of appearance and
%disappearance channels (red solid curve in the figure) 
%gives increased values of $\chi^2_{min}$ compared to the individual channels.

\item From
Figure~\ref{NovaT2K_testth23} (right panel) , we see that the $\chi^2$ for the
disappearance channel is weakly dependent on the test value of $\dcp$,
and remains small for all values of test $\dcp$. 
%, since it is the
%marginaliz ation over $\theta_{23}^{tes t}$ which has a major
%contribution for this channel and the effect of tes t $\dcp$ is
%negligible.
On the other hand, the $\chi^2$ for the appearance channel
has a strong dependence on test $\dcp$, which is consistent with the
behaviour of the respective probabilities seen in Section~\ref{subsec:prob}.
%Therefore, the combined $\chi^2$ from the appearance and disappearance
%channels shows a significant drop with marginalization over test
%$\dcp$.
In terms of test $\dcp$, the minima of
the combination occur at test $\dcp$ values close to those for the
appearance channel, which is the principal contributor to the $\dcp$
dependence of the sensitivity. 
The increased values of the combined \nova\ $\chi^2$ minima 
at each test $\dcp$, 
can be attributed to the 
tension  
between the two channels with respect to $\theta_{23}^{test}$.
Thus marginalization over $\dcp$ reduces the sensitivity to some extent 
because of
the effect of $\dcp$ uncertainty in the appearance channel.

%\item There is also a tension between the location of $\chi^2$ minima as a
%function of test $\delta_{CP}$; for example, for $\theta_{23}^{tr} = 36^o$, $\chi^2_{min}$ is
%at $\delta_{CP} = 0$ for the appearance channel and at $\delta_{CP} = \pi$ for the
%disappearance channel. {\bf{The $\chi^2$ is seen to be very weakly dependent on
%the test value of $\delta_{CP}$ for the disappearance channel}}.

\item The $\chi^2$ minima for both appearance and disappearance channels are
very low for all values of $\theta_{23}^{tr}$, indicating negligible
octant sensitivity 
from the channels separately. This is apparent from the left panel of Figure 
\ref{NovaT2K_testth23}, where the minimum values of $\chi^2$ for individual channels 
are quite small when marginalized over the entire $\theta_{23}$ range. 

%\item
%In the combination of NOvA appearance and disappearance channels,
%the $\chi^2$ minima are near $\theta_{23}^{test} = \pi/2 - \theta_{23}^{tr}$,
%following
%the behaviour of the disappearance channel $\chi^2$, but the values are enhanced
%due to the conribution from the appearance $\chi^2$ values at that point.
%In terms of $\delta_{CP}$, the minima of the combination occur at test $\delta_{CP}$
%values in between those for the individual channels. Due to both these
%synergies, the combined NOvA $\chi^2$ minima have increased values
%indicating non-negligible octant sensitivity. 

%{\bf{The values are asymmetric across the lower and higher octants and strongly dependent
%on the true value of $\delta_{CP}$. For example, $\chi^2_{min}$ is about 1.5 for NOvA (appearance $+$ disappearance) at 
%$\theta_{23}^{tr} = 36^o$ with $\sin^2 2\theta_{13}^{tr} = 0.1, \delta_{CP}^{tr} = 0$. For $\theta_{23}^tr = 54^o$, $\chi^2_{min}$ goes 
%up to about 11 for NOvA (app $+$ disapp) for  the same value of $\delta_{CP}^{tr}$. For $\delta_{CP}^{tr} = \pi/2$, 
%$\chi^2_{min}$ for NOvA (app $+$ disapp) is 5.6 for $\theta_{23}^{tr} = 36^o$ and 7.5 for $\theta_{23}^{tr} = 54^o$}}.

\item There is no major tension between \nova\ appearance $+$
disappearance and T2K
appearance $+$ disappearance. The $\chi^2$ minima occur at the same values of test
$\sin^2 2\theta_{13}$ and test $\theta_{23}$, and for somewhat displaced values of test
$\delta_{CP}$. This is because both T2K and \nova\ roughly follow the octant
behaviour dictated by the vacuum probabilities with subleading
contributions of the $\delta_{CP}$-dependent terms. The slightly increased
matter effect contribution for \nova\ does not seem to have a
significant effect. 
%Because of the slight tension between the $\delta_{CP}$ 
%dependence of NOvA and T2K, the effect of $\delta_{CP}$ uncertainty is somewhat 
%reduced by the combination. 
The combined \nova$+$T2K $\chi^2$ values are 
nearly equal to the sum of the corresponding individual \nova\ and
T2K values, as seen in Table~\ref{tablelimits_NovaT2K}.{\footnote{{In this table
and some subsequent figures and tables, the $\chi^2$ values without priors are given 
only for comparison of the relative weightage of the contributions
from different channels and different experiments.}}}
For example, $\chi^2_{min}$ for T2K appearance $+$ disappearance is about 3 for
$\theta_{23}^{tr} = 54^o$. The corresponding \nova\ value is about 11. For
\nova$+$T2K, $\chi^2_{min}$ is about 15 for the same $\theta_{23}^{tr}$.

\item The addition of priors to \nova$+$T2K  enhances the sensitivity
drastically,  
making it (for example) $(\chi^2_{min})^{prior} = 21$ for $\theta_{23}^{tr} = 54^o$
{{with the present value of the $\theta_{13}$ prior}}.
This underscores the importance 
of precision measurement of $\theta_{13}$ on octant 
sensitivity. 
With more precise measurements of $\sin^2 2\theta_{13}$ expected from the 
reactor experiments, 
the octant sensitivity is expected to get better. 
{{If we take a stronger projected prior of $\sigma_{\sin^2 2\theta_{13}} = 0.005$ instead of the present value of 
 $\sigma_{\sin^2 2\theta_{13}} = 0.01$, the results improve further, as seen in column 5 of the table, where ${\rm{prior_{n}}}$ 
denotes the projected prior. Here, for example, the $\chi^2$ goes up from $3.9$ to $6.3$ for 
$\theta_{23}^{tr}=39^\circ$, $\dcp^{tr}=0$ and normal hierarchy.
In the limit of infinite precision, 
i.e. if $\theta_{13}$ is held fixed, the $\chi^2$ goes up further to $7.7$ for 
the same case}}.

\item 
We see from the 
Figures~\ref{octant_NovaT2K}
and \ref{octant_NovaT2K_marghier} that 
the $\chi^2$ is asymmetric between the lower and higher octants.
The nature of this asymmetry depends on the 
true $\delta_{CP}$ value. For instance for $\delta_{CP}^{tr} = 0$,             
the $\chi^2$ values on the LO side are much lower than on
the HO side, but this is reversed for other $\delta_{CP}^{tr}$ values.
Note that the addition of priors makes 
the asymmetry less pronounced since
the prior contributions for the lower and higher octants are roughly 
symmetrical. This can be seen by comparing the with and without prior 
values of $\chi^2$  in 
Table~\ref{tablelimits_NovaT2K}.

\end{enumerate}

%\section{Experimental specifications}
\subsection{Atmospheric neutrinos} 

For our study of octant sensitivity in atmospheric neutrino experiments, 
we look at the following set-ups:
\begin{enumerate}
\item A large magnetized iron detector  with an exposure of 
500 kT yr, capable of detecting muon events with charge 
identification, using the following neutrino energy and 
angular resolution:
$\sigma_{E_{\nu}} = 0.1 \sqrt{E_{\nu}}$, $\sigma_{\theta_{\nu}} = 10^o$.

%For the atmospheric analysis, the flux and detector systematic uncertainties
%are included using the method of $\chi^2$ pulls as outlined in \cite{gandhi}.
Note that such a detector will be constructed by the India-based Neutrino 
Observatory (INO) collaboration~\cite{ino}. 
The energy and angular resolutions of  muons are available from 
INO simulation code~\cite{gct,choubey-nu2012}.
But the work on reconstruction of neutrino energy 
and angle requiring the resolutions for  both muons and hadrons
are in progress. In our neutrino analysis therefore we use 
fixed resolutions in terms of neutrino energy and angle
as described above. 
Determination of the octant sensitivity of INO using the 
resolutions obtained from INO simulations is in progress.

\item A LArTPC with an exposure of 500 kT yr (unless otherwise stated)
capable of detecting muon and electron events. No charge 
identification is assumed here. The angular resolutions are taken to be 
\cite{ourprl}: 

\begin{eqnarray}
\sigma_{\theta_{\nu e}} = 2.8^o, \nonumber \\
\sigma_{\theta_{\nu \mu}} = 3.2^o
\end{eqnarray}

For the neutrino energy resolutions, we use the estimated value 
$\sigma_{E_{\nu}} = 0.1 \sqrt{E_{\nu}}$. A LArTPC has a very good 
angular resolution since ionization tracks
can be transported undistorted over distances of several metres in highly 
purified LAr, allowing for excellent
direction reconstruction by recording several projective views of 
the same event using wire planes with different
orientations~\cite{larmag0}.
\end{enumerate}

\subsection{Octant sensitivity using atmospheric muon events in a 
magnetized iron detector}

\FIGURE{
\includegraphics[width =\textwidth,clip=,angle=0]{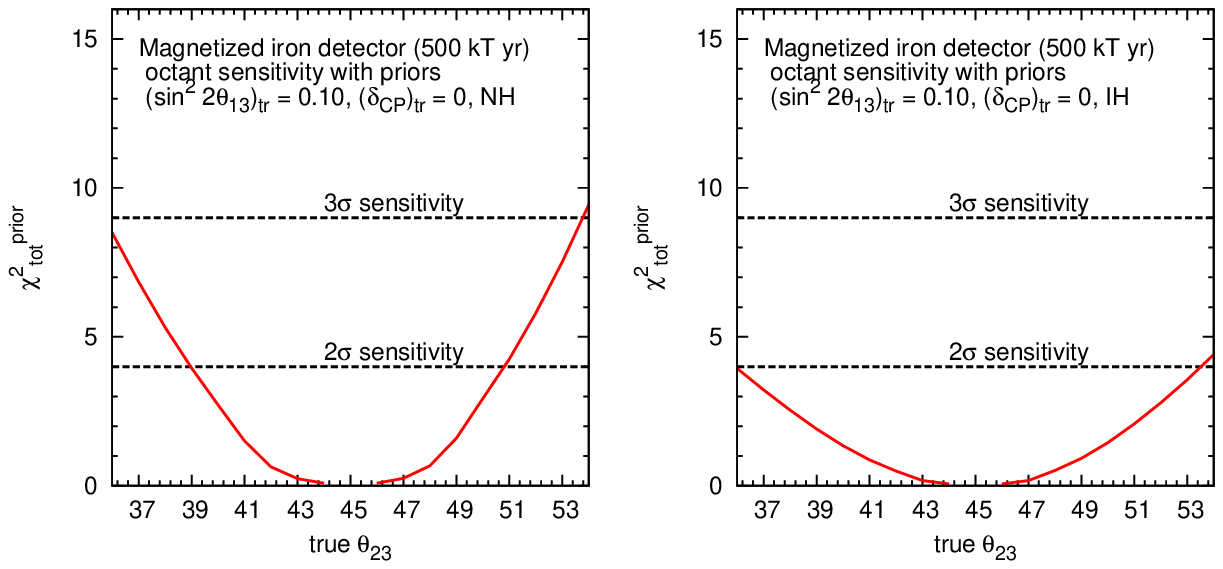}
\vspace{-0.4in}
\caption{\small Marginalized octant sensitivity from muon 
events with a magnetized iron detector (500 kT yr exposure), 
for the case of normal and inverted mass hierarchy. In this figure 
priors have been added.}
\label{octant_ICAL}
}

The $\chi^2$ is defined as, 
\bea
{\rm{(\chi^2_{tot})^{prior}}} = {\rm{min(\chi^2_{Atm} + \chi^2_{prior})
}} ~,
\eea
where,
\begin{equation*}
 \textrm{where  } \rm{\chi^2_{Atm}} =  
 \chi^2_\mu + \chi^2_{\bar{\mu}}  
\end{equation*}
For the atmospheric analysis, the flux and detector systematic uncertainties
are included using the method of $\chi^2$ pulls as outlined in 
\cite{gandhi,pulls,pulls2,pulls3}.

It is well known that the possibility of magnetization 
gives the iron detectors an excellent sensitivity to hierarchy 
\cite{hierical_indu,hierical_petcov,gandhi,hierical_samanta,hierical_blennow,gct}. 
Therefore it is plausible to 
assume that hierarchy would  be determined before octant  in 
these detectors. 
Thus in our analysis of atmospheric neutrinos, we assume  
the neutrino mass hierarchy fixed to be either normal or inverted,
and do not marginalize over the hierarchy. 
This is tantamount to the  assumption that
the wrong-octant-wrong hierarchy
solutions are excluded.  
In Figure~\ref{octant_ICAL}, we present the octant sensitivity of a 
magnetized 
iron calorimeter detector with an exposure of 500 kT yr. 
The second column of Table~\ref{chisqfinal} displays the 
$\chi^2$  values for different values of $\theta_{23}$ in the LO. 
The $\chi^2$ in this case is symmetric about $\pi/4$, and therefore the 
values in the higher octant are similar.  
Assuming an inverted hierarchy gives worse results because 
in this case, matter resonance and hence octant sensitivity occurs 
in the antineutrino component of the event spectrum, for which 
the flux and detection cross-sections are lower.   
A 2$\sigma$ sensitivity is obtained at $\theta_{23} = 39^o$ if NH 
is the true hierarchy. 
 
\subsection{Octant sensitivity using atmospheric events in a LArTPC}
%\subsection{Using atmospheric muon or electron events individually (Liquid Argon detector)} 

A LArTPC is sensitive to both muon and electron type events 
and one can study the interplay of both type of events in  
octant sensitivity using atmospheric neutrinos. 
We give the results separately for the atmospheric muon and electron signals 
as well as for the combined  analysis.

In the following          
analysis we assume
that the hierarchy will be determined 
by the combination of reactor, long baseline  and INO experiments 
\cite{gct} before a LArTPC can give an octant measurement. 
Therefore we do not marginalize over hierarchy in the wrong octant. 

The muon and electron event spectra are related to the oscillation probabilities as follows:
\begin{eqnarray}
N_{\mu} \sim (\phi_{\mu} P_{\mu\mu} + \phi_{e} P_{e \mu}) \times {\rm{(CC~cross-section, exposure, efficiency)}} \nonumber \\
N_{e} \sim (\phi_{\mu} P_{\mu e} + \phi_{e} P_{ee}) \times {\rm{(CC~cross-section, exposure, efficiency)}} 
\end{eqnarray}

%WHY ``+C.C.'' IN THE FORMULA FOR EVENT RATES? WE ARE ALSO DISCUSSING RESULTS WITH 
%CHARGE ID. SHOULD WE MAKE THIS DISCUSSION GENERAL BY REMOVING ``+C.C.''?
%\\

The antineutrino event rates are given by the same expressions 
with the flux and probabilities replaced by their antineutrino
counterparts.

The $\chi^2$ in this case is given by
\bea
{\rm{(\chi^2_{tot})^{prior}}} = {\rm{min(\chi^2_{Atm} + \chi^2_{prior})
}} ~,
\eea
where,
\begin{equation*}
 \textrm{where  } \rm{\chi^2_{Atm}} =  
 \chi^2_{\mu+{\bar{\mu}}} + \chi^2_{e+\bar{e}}  
\end{equation*}

The atmospheric muon flux is approximately twice that of the electron flux. 
Hence the behaviour of the muon events is dictated by the muon survival
probability 
and to a lesser extent by the $\pemu$ oscillation probability. 
The sensitivity derivable 
from muon events is
somewhat compromised by the fact that $\pemu$ and $\pmumu$ shift in opposite 
directions for a transition from
one octant to another, as observed in Figure~\ref{PvsE_th13}. On the other 
hand, for electron events the only source of the octant sensitivity is
from $\pmue$, since the electron survival probability is 
independent of $\theta_{23}$. Therefore both muon and electron events
individually give comparable values of octant sensitivity. 
The sensitivity from muon events is more strongly dependent on the value of 
$\theta_{23}^{tr}$, and is therefore higher than the sensitivity from electron events
for values further away from maximal, getting closer to the latter as $\theta_{23}^{tr}$ 
approaches $45^o$. This can be seen from 
Table~\ref{tablelimits} 
%and \ref{tablelimits_IH}
where 
we  list the values of octant sensitivity 
from atmospheric muon and electron events
in a LArTPC with an exposure of 500 kT yr, marginalized over
the allowed ranges of the oscillation parameters. The neutrino mass hierarchy is fixed 
to be either normal or inverted.
%assuming that the hierarchy would be determined before the octant in these type of 
%detectors \cite{ourprl}.
The results presented are for true values
of $\theta_{23}$ lying in the lower octant. The behaviour for $\theta_{23}^{tr}$ in 
the upper octant is similar
and symmetrical about $\theta_{23}^{tr} = 45^o$. 

\begin{table}
%\TABLE[!h]{\centerline{
\begin{center}
%\TABULAR
\begin{tabular}{|c || c | c | c|} \hline
        {{$\theta_{23}^{tr}$}} & {{$\chi^2_{\mu}$}}
& {{$\chi^2_{e}$}} &  {{$\chi^2_{\mu+e}$}} \\
& NH(IH) & NH(IH) & NH(IH) 
        \\
        \hline
    36 & 8.5 (1.5)& 3.5 (1.4)& 15.4 (5.8)\\ \hline
     37 & 6.7 (1.3)& 2.8 (1.1)& 11.5 (4.8)\\ \hline
     38 & 4.6 (1.0)& 2.5 (0.8)& 9.0 (3.7)\\ \hline
        39  &  3.0 (0.7)& 1.7 (0.6)& 7.1 (2.8)\\ \hline
         40 & 1.9 (0.6)& 1.4 (0.5)& 4.8 (1.7)\\ \hline
  41 & 1.0 (0.3)& 1.0 (0.3)& 3.0 (0.9)   \\ \hline
  42 & 0.6 (0.2)& 0.6 (0.2)& 1.9 (0.5)  \\ \hline 
  43 & 0.3 (0.1)& 0.4 (0.1)& 1.0 (0.3)  \\ \hline
     44  & 0.1 (0.1)&  0.2 (0.1)& 0.4 (0.2) \\ \hline
%        0.12  & 15.5 &  8.3 & 12.1  
%\\ \hline
%        0.15  & 6.1 & 7.1 & 11.8 & 16.9        

%       $\sin^2 \theta_{23} $ & 0.50 &  0.34 -- 0.68 \\ \hline
%       $\sin^2 \theta_{13} $ & 0.00 &  $\le$ 0.047
           \hline
\end{tabular}
\caption{\small Marginalized octant sensitivity from atmospheric muon, electron 
and muon $+$ electron events 
with a LArTPC (500 kT yr exposure), for the case 
of normal (inverted)
mass hierarchy, with $\sin^2 2\theta_{13}^{tr} = 0.1$ and $\dcp^{tr} = 0$. 
Priors have not been included here.}
\label{tablelimits}
\end{center}
\end{table}

\begin{figure*}[t]
\includegraphics[width =\textwidth,clip=,angle=0]{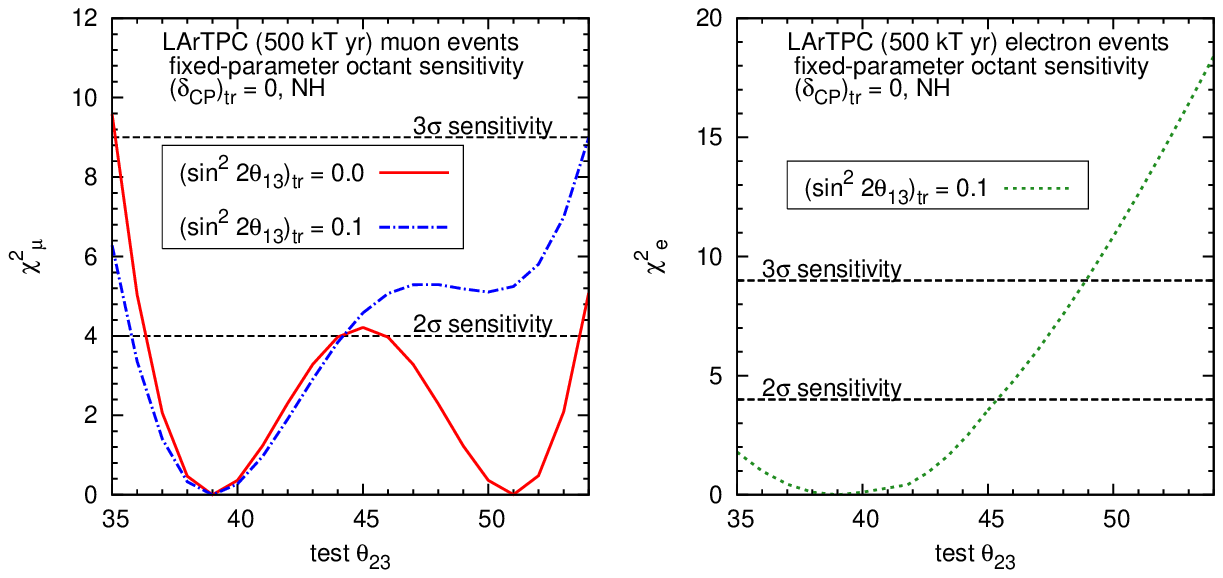} 
\vspace{-0.3in}
\caption{{\small{Fixed-parameter octant sensitivity from atmospheric neutrinos with 
a LArTPC (500 kT yr exposure) as a function of test $\theta_{23}$
for $(\sin^2 2 \theta_{13})_{tr} = 0.1$ and $(\theta_{23})_{tr} = 39^o$. The left 
panel is for muon events and the right panel is for electron events. 
 }}}
\label{octant_testth23}
\end{figure*}

Figure~\ref{octant_testth23} shows the behaviour of the octant sensitivity with 
the atmospheric neutrino signal in a LArTPC for a 
specific true value $(\theta_{23})^{tr} = 39^o$, as a function of 
the test values of $\theta_{23}$, for both muon  (left panel) 
and electron  (right panel)  events.
For the  latter the only octant-sensitive contribution 
is from $\pmue$, which does not suffer from the
intrinsic octant degeneracy 
at the leading order and the octant sensitivity is due to the
$\sin^2 \theta_{23}  \times \sin^2 2\theta_{13}^m$
term.
Since $\sin^2 2\theta_{13}^m$ becomes close to
1 near matter resonance, 
the octant sensitivity increases proportionally with the test value of 
$\sin^2 \theta_{23}$.
Hence the minimum sensitivity is seen in the region near
$\theta_{23}^{test} \sim 45^o$ in the wrong octant.
The solid red curve in the left panel
is for $\sin^2 2\theta_{13} = 0$ for which the 
contribution comes from $\pmumu$ only. 
The curve exhibits 
no octant sensitivity   
due to the 
intrinsic degeneracy coming from the $\sin^2 2\theta_{23}$ term 
in $\pmumu$. 
This degeneracy is broken for higher 
values of $\sin^2 2\theta_{13}$ where both $\pmumu$ and $\pemu$
can contribute. 
%For $\pmumu$ ($\pemu$)  this is due to the 
%the octant sensitive term $\sin^4 \theta_{23} (\sin^2 \theta_{23})
%\times \sin^2 2\theta_{13}^m$. 
As a combined effect of all these factors  
%the  
%minimum separation between octants is no longer 
%located near $\theta_{23}^{test} = 90^o - \theta_{23}^{tr}$. 
%Hence 
the octant sensitivity does not have a clean proportionality with 
$\theta_{23}^{test}$, but can have a minimum anywhere within the range 
$45^o < \theta_{23}^{test} < (90^o - \theta_{23}^{tr})$
depending on the relative weightage of the two terms
for specific energies, baselines
and values of the neutrino parameters.
This is reflected by the blue dot-dashed 
curve  in the left panel. 

%\subsection{Combining atmospheric electron and muon events (Liquid Argon detector)} 
%As can be seen from the previous discussion, the octant sensitivities 
%using the muon and electron event spectra individually are quite low.  
We now study whether combining muon and electron events will give 
an enhanced octant sensitivity compared to that from the muon and electron 
event spectra individually. The last column of 
Table~\ref{tablelimits} lists  the 
$\chi^2_{\mu+e}$ values for NH (and IH). These values are obtained 
by marginalizing over all  test parameters, but without adding priors.    
%Here we give the combined marginalized octant sensitivity that can 
%be derived using both muon and electron events, listed in Table 4. 
In Figure~\ref{octant_LAr} the sensitivity is presented with the
inclusion 
of priors. 
The following points may be noted from the listed results:
\begin{enumerate}
\item The values of the octant sensitivity that can be derived 
from the atmospheric muon and electron events separately 
are relatively low.
% compared to that from a combined analysis, 
%reaching a level of $ \sim 3\sigma$ for  
%$\theta_{23}^{tr} = 36^o$ with muon events and an assumed normal 
%hierarchy for an exposure of 500 kT yr. 
A combined analysis of muon and electron events gives improved 
sensitivities which are better than the sum of the marginalized 
$\chi^2$ sensitivities from the two kinds of signals. This is 
because the muon and electron event spectra behave differently as 
functions of $\theta_{23}$ and hence their $\chi^2$ 
minima occur at different parameter values. Utilizing this fact, 
a marginalization over the sum of $\chi^2$ values gives an 
improved result compared to the individual contributions. The enhancement
in the sensitivity due to this synergy can be anywhere between 20$\%$ and 50$\%$, depending on
the true value of $\theta_{23}$, as can be seen in Table~\ref{tablelimits}.
%and \ref{tablelimits_IH}. 

\item The addition of prior information further improves the 
sensitivity by about 20 - 40$\%$. The most significant contribution comes from the 
reduced error range of the parameter $\theta_{13}$ from recent 
reactor results. For an assumed normal hierarchy, a 3$\sigma$ 
signal of the octant may be achieved for $\theta_{23}^{tr} = 39^o (51^o)$ 
for a true lower octant (higher octant) with this exposure, as seen in Figure~\ref{octant_LAr}.  

\item Assuming an inverted hierarchy gives worse results because 
in this case, matter resonance and hence octant sensitivity occurs 
in the antineutrino component of the event spectrum, for which 
the detection cross-sections are lower.   
\end{enumerate}

\FIGURE{
\includegraphics[width =\textwidth,clip=,angle=0]{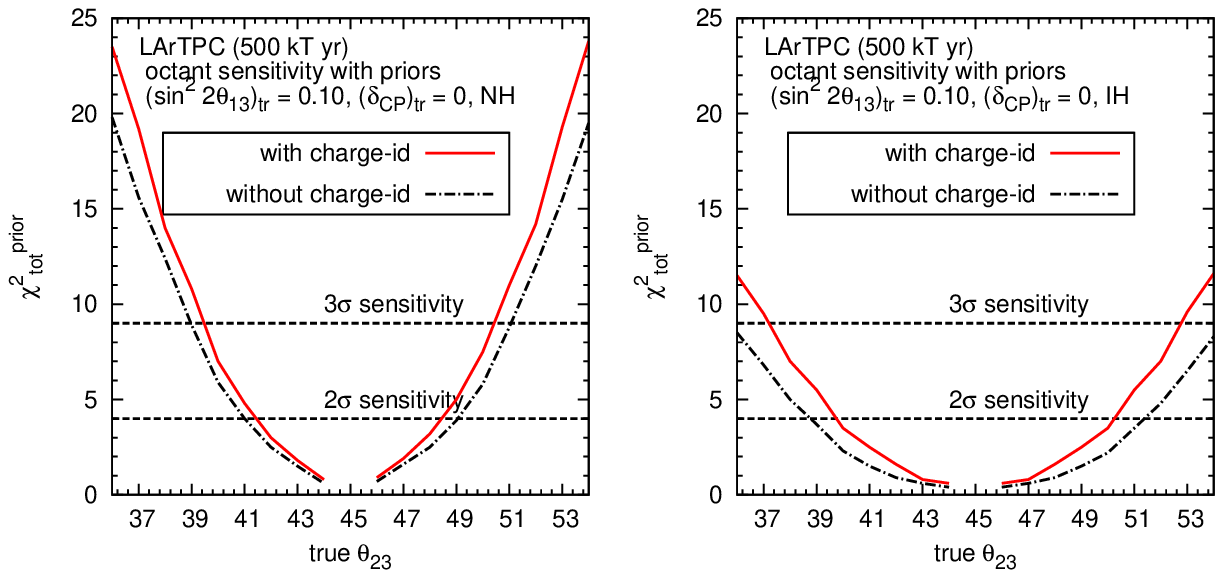}
\vspace{-0.4in}
\caption{\small Marginalized octant sensitivity from both muon 
and electron events with a LArTPC (500 kT yr exposure), 
for the case of normal and inverted mass hierarchy. In this figure 
priors have been added.}
\label{octant_LAr}
}

\FIGURE{
%\vskip 1cm
%\includegraphics[scale=.55]{}
%\includegraphics[scale=.6]{prob1}
\includegraphics[width=0.7\textwidth]{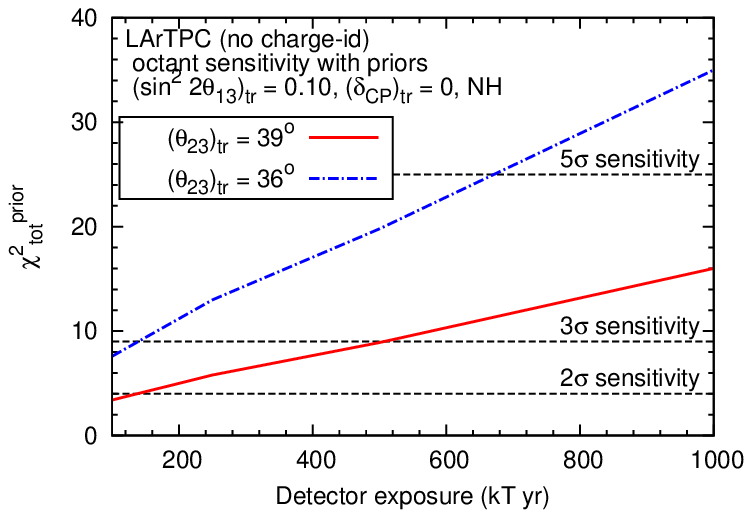}
\caption{{\small {Octant sensitivity for a non-magnetized LArTPC as 
a function of exposure. $\chi^2$ has been plotted for normal hierarchy for two values of 
$\theta_{23}$(true) -- $36^\circ$ and $39^\circ$. 
 }}}
\label{octant_exposure}
}

In Figure~\ref{octant_exposure}, the marginalized octant sensitivity from a LArTPC 
is plotted as a function of detector exposure for two true values of $\theta_{23}$. 
The figure shows that it is possible to achieve a 2$\sigma$ sensitivity 
for $\theta_{23}^{tr} = 39^o$
and a 3$\sigma$ sensitivity for $\theta_{23}^{tr} = 36^o$ from a LArTPC alone
for an exposure as low as 120 kT yr, in the case of a normal hierarchy.

\subsection{Effect of magnetization}
Figure~\ref{octant_LAr}  shows the 
values 
of octant sensitivity derivable from a LArTPC with and 
without magnetization.  
The magnetization offers the possibility of charge identification.  
In the case of a LArTPC with magnetization, charge identification details
are incorporated as given in Refs.~\cite{ourprl,larmag1,larmag2}, assuming 
a 100$\%$ charge identification capability
for muon events and a 20$\%$ charge identification capability in the 
energy range 1-5 GeV (none for higher energies) for electron events.

The  octant sensitivity without charge identification  
is seen to be about 80$\%$ (60$\%$) of that with 
charge identification capability 
with a similar exposure,
in the case of a normal (inverted) mass hierarchy. 
This 
difference in behaviour is explained as follows. 

If the true hierarchy is normal, neutrino events have octant sensitivity due to resonant 
matter effects, while antineutrino events do not. Therefore the antineutrino events for 
both octants are almost the same, say $k$. In the absence of 
charge-identification, we add these events, and 
\[
\chi^2 = \frac{ ( [N^{HO}+k] - [N^{LO}+k] )^2 }{ N^{HO}+k } . 
\]
On the other hand, if we do have charge-identification, then the chisq is simply 
\[
\chi^2 = \frac{ ( [N^{HO}] - [N^{LO}] )^2 }{ N^{HO} } + \frac{ ( [k] - [k] )^2 }{ k }. 
\]
In both cases, the numerator is same, but the denominator is more in the former
case. Therefore charge-identification gives us higher sensitivity. 

Note that when the hierarchy is normal, $k$ comes from the antineutrino events. 
Because of the flux and cross-section being small, $k$ is a relatively small number. 
But if the hierarchy is inverted, then $k$ will be because of neutrino events, whose 
higher flux and cross-section will make the 
denominator quite large. Therefore, the reduction in sensitivity due to loss of 
charge-identification will be more apparent for the inverted hierarchy.
Charge identification capability has been shown to play 
a very crucial role in determination of mass hierarchy in such detectors 
\cite{ourprl}. 
However this does not seem to play such an important role for octant 
determination if we already assume a prior knowledge of the hierarchy. 
{{For a LArTPC without charge indentification, if we assume that the hierarchy is not known then the octant sensitivity is unaffected
for $|\theta_{23}^{tr} - 45^o| < 7^o$, i.e. for $39^o \leq \theta_{23}^{tr} \leq 51^o$, but for smaller (larger)
values of $\theta_{23}^{tr}$ in the lower (higher) octant, there is a drop of 
20-30$\%$ 
in the sensitivity.
For a magnetized iron detector,
we have checked that the effect of marginalizing over the hierarchy is negligible since 
the wrong hierarchy is excluded with a reliable confidence
level for the exposure considered (500 kT yr). The same is true for a LArTPC with charge identification capability}}.

When we compare the octant sensitivity 
from a magnetized iron detector with the LArTPC results without 
charge identification, the 
iron detector sensitivities are about 40$\%$ of those with a LArTPC 
if the hierarchy is normal and about 50$\%$ of the 
LArTPC sensitivities if the hierarchy is inverted.
The sensitivities from a LArTPC are higher than those from 
a magnetized iron detector . This is because of the possibility of realizing 
very high angular resolutions in the former type of  detector and
the significant contributions from muon as well as electron events.

\subsection{Effect of $\dcp$}

The dependence of the octant sensitivity on the CP phase $\dcp$ is different for
atmospheric neutrino experiments and the long baseline experiments considered.
Because of the strong earth matter effects over a large range of atmospheric 
neutrino baselines,
the behaviour of the corresponding oscillation probabilities is governed by 
the enhanced resonant
features and their dependence on $\dcp$ is suppressed. On the other hand, 
for \nova/T2K baselines, since the matter effects are smaller, $\dcp$ plays a greater role
in the probability behaviour and hence in the octant sensitivity, because of the degeneracy 
of the octant with $\theta_{13}$ and $\dcp$ as explained previously. 

\FIGURE{
\epsfig{file=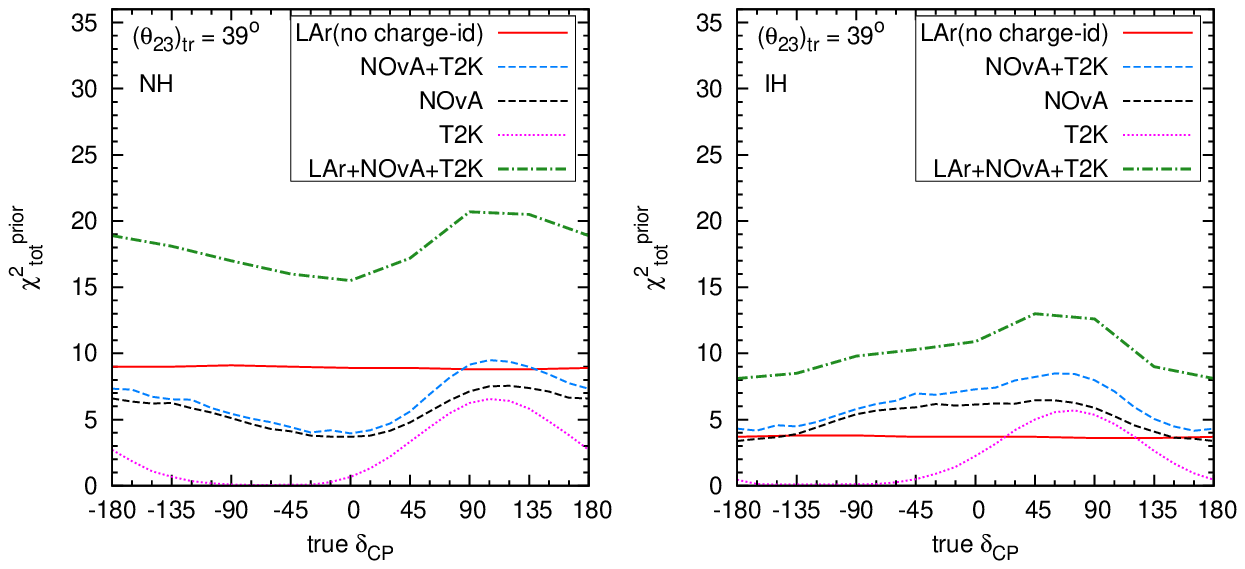,width=\textwidth} \\
\epsfig{file=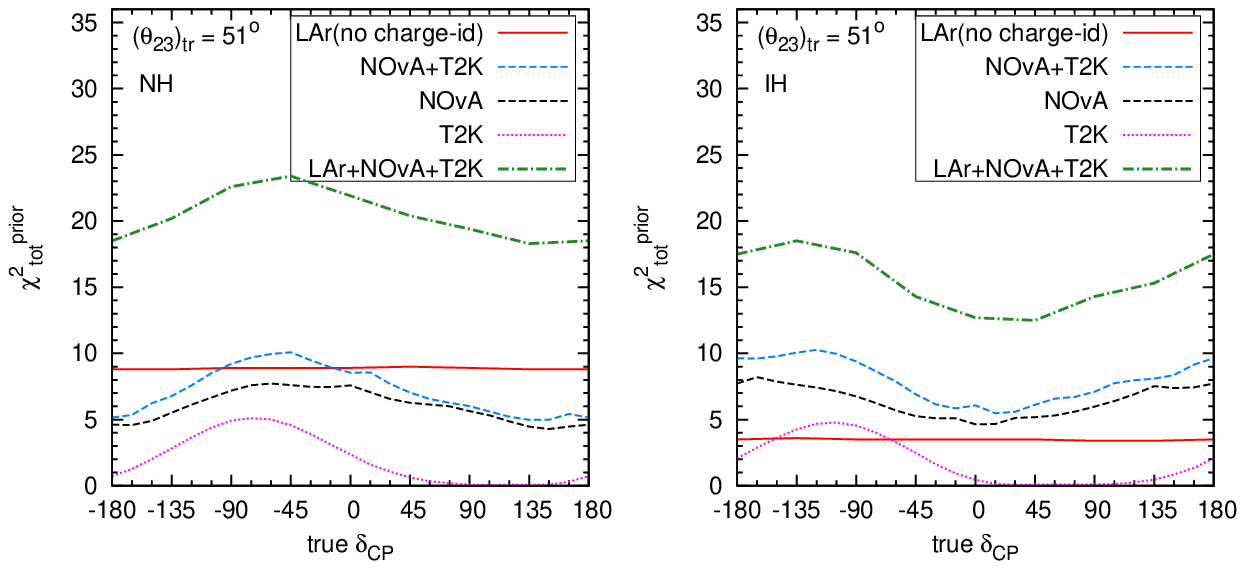,width=\textwidth} \\
\vspace{-0.4in}
\caption{\small Marginalized octant sensitivity with priors as a function of true 
$\dcp$ from \nova, T2K, \nova\ $+$ T2K and for an atmospheric neutrino experiment with 
a LArTPC for the case of normal and 
inverted mass hierarchy, for 2 values of $\theta_{23}^{tr}$ in the lower and
higher octants.}  
\label{octant_dcpt23}
}

\begin{figure*}[t]
\hspace*{-0.6in}
\includegraphics[scale=1.2]{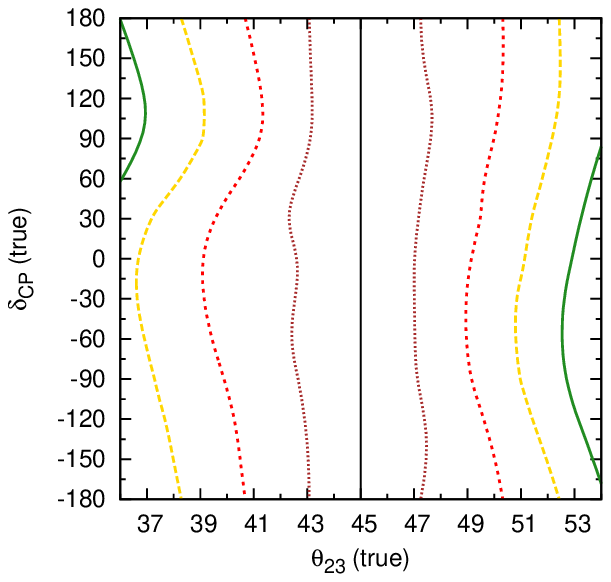}
\hspace*{-1.2in}
\includegraphics[scale=1.2]{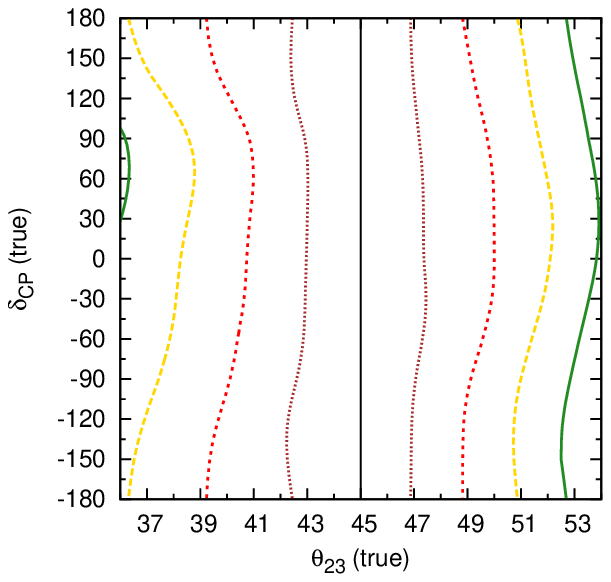}
\vspace{-0.3in}
\caption{\small {Wrong octant exclusion for \nova+T2K in the true $\theta_{23}$ - true 
$\dcp$ plane. The left(right) panel is for NH(IH) as the true hierarchy. The 
brown(dense-dotted)/red(sparse-dotted)/gold(dashed)/green(solid) contours denote 
$1/2/3/4 \sigma$ exclusion, respectively.
}}
\label{octant_contdcpt23}
\end{figure*}

%In Figure \ref{octant_dcpt23} the fixed parameter octant sensitivity is depicted 
%in the form of contours
%in the true $\theta_{23}$ - true $\dcp$ plane for $(\sin^2 2 \theta_{13})_{tr} = 0.1$.
%The left panel gives the sensitivity derived from NOvA/T2K, and shows that there is
%some dependence on the value of $\dcp$. The right panel, giving the atmospheric neutrino
%octant sensitivity with a Liquid Argon detector, consists of contours which are almost parallel
%to the $\dcp$ axis, confirming that the effect of $\dcp$ on the octant sensitivity from this source is insignificant.  

These features are reflected 
in Figure~\ref{octant_dcpt23}, in which the marginalized octant sensitivity for a LArTPC 
with atmospheric neutrinos, \nova, T2K and their combinations are plotted as 
a function of the true $\dcp$ for $\theta_{23}^{tr} = 39^o$ (LO) and $51^o$ (HO)  
for both normal and inverted mass hierarchies.
% The figure shows the strong dependence 
%of the \nova and T2K sensitivities on the true value of $\dcp$, while for a LArTPC 
%the effect of $\dcp$ is insignificant.
The $\dcp$ behaviour of the 
\nova\ and T2K sensitivities are seen to be similar, and they follow an opposite 
behaviour  with respect to  $\dcp$ 
for $\theta_{23}^{tr}$ lying in the lower and higher octants,
i.e. for $\theta_{23}^{tr} = 39^o$ they are higher in the range $0^o < \dcp^{tr} < 180^o$ 
and for $\theta_{23}^{tr} = 51^o$ they are higher in the range $-180^o < \dcp^{tr} < 0^o$
\cite{uss}. 

For $\theta_{23}^{tr}= 39^o$  (LO) and NH (the top left panel) the contribution from a
LArTPC is higher than that from combined \nova+T2K excepting for a narrow 
range $\dcp^{tr} = 90^o$ to $135^o$. 
For  $\theta_{23}^{tr} = 51^o$ (HO) and NH (bottom left panel) 
the same trend is observed, except that in this case the combined \nova+T2K  has a higher 
sensitivity for the range $\dcp^{tr} = -15^o$ to $-90^o$.

For the top (bottom) right panels which are for $\theta_{23}^{tr}= 39^o$ ($51^o$) and IH, the 
\nova\ $+$ T2K as 
well as the standalone \nova\ contribution to the sensitivity are higher than the 
atmospheric contribution over almost all values of true $\dcp$. 
This is because the sensitivity from atmospheric neutrinos is
less than half for an inverted hierarchy compared to normal hierarchy, while
for \nova\ and T2K the values are similar.

The T2K sensitivity is lower than \nova\ in all cases, and lower than
the atmospheric contribution for most values of true $\dcp$, besides certain small 
ranges in the inverted hierarchy case where it is higher.  The combined sensitivity 
from atmospheric neutrinos and \nova\ $+$ T2K follows the true $\dcp$ dependence 
of the \nova\ $+$ T2K sensitivity.

Finally, in Figure~\ref{octant_contdcpt23}, we show the wrong octant exclusion sensitivity 
for \nova+T2K in the true $\theta_{23}$ - true $\dcp$ plane. Contours are shown in this 
plane for $1$, $2$, $3$ and $4 \sigma$ octant exclusion. In the part of the parameter space 
that is enclosed within a certain contour, it is possible to exclude the wrong octant with 
the corresponding confidence level. As expected, the ability of 
the experiments to exclude the wrong octant is small around 
true $\theta_{23} \simeq 45^\circ$. As $\theta_{23}$ becomes more non-maximal, the exclusion 
ability becomes better. The effect of $\dcp$ on octant sensitivity for any given true value 
of $\theta_{23}$ can also be read from these plots. Since the results using atmospheric 
neutrinos are almost independent of $\dcp$, the behaviour of the $\chi^2$
is similar to the 
plots in Figures~\ref{octant_ICAL} and 
\ref{octant_LAr} drawn for $\dcp=0$.

\subsection{Octant sensitivity  from combined analysis of
atmospheric electron and muon events with \nova\ and T2K}

\FIGURE{
\includegraphics[width =\textwidth,clip=,angle=0]{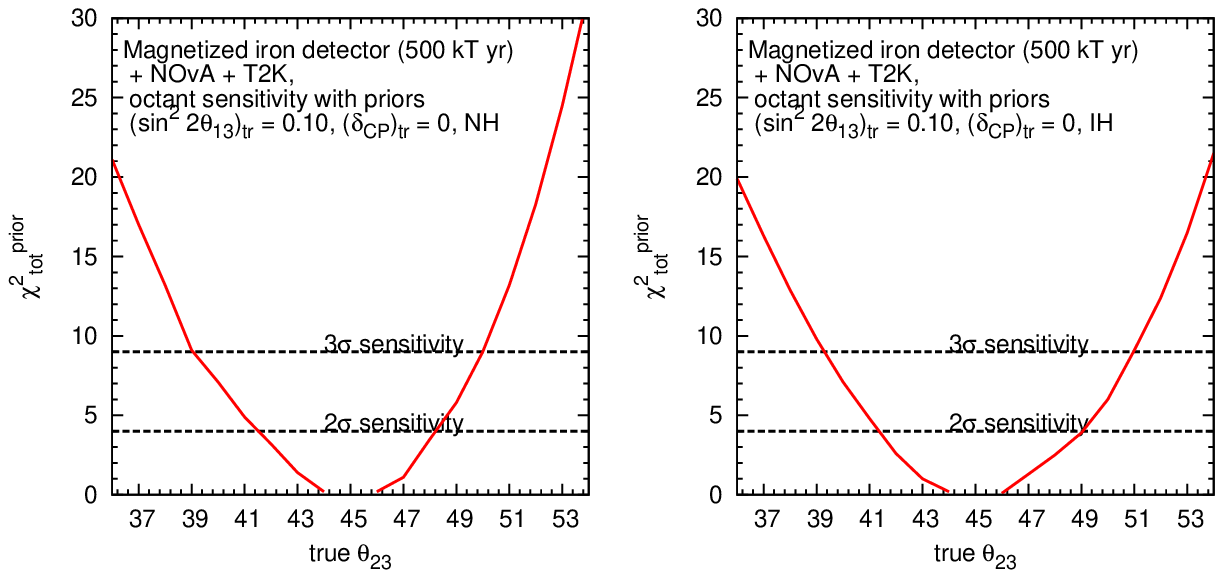}
\vspace{-0.4in}
\caption{\small Marginalized octant sensitivity from a combination of the atmospheric 
muon neutrino signal in a magnetized iron calorimeter detector (500 kT yr) $+$ \nova\ $+$ 
T2K, for the case of normal and inverted mass hierarchy. In this figure 
priors have been added.}
\label{octant_ICALNovaT2K}
}

\FIGURE{
\includegraphics[width =\textwidth,clip=,angle=0]{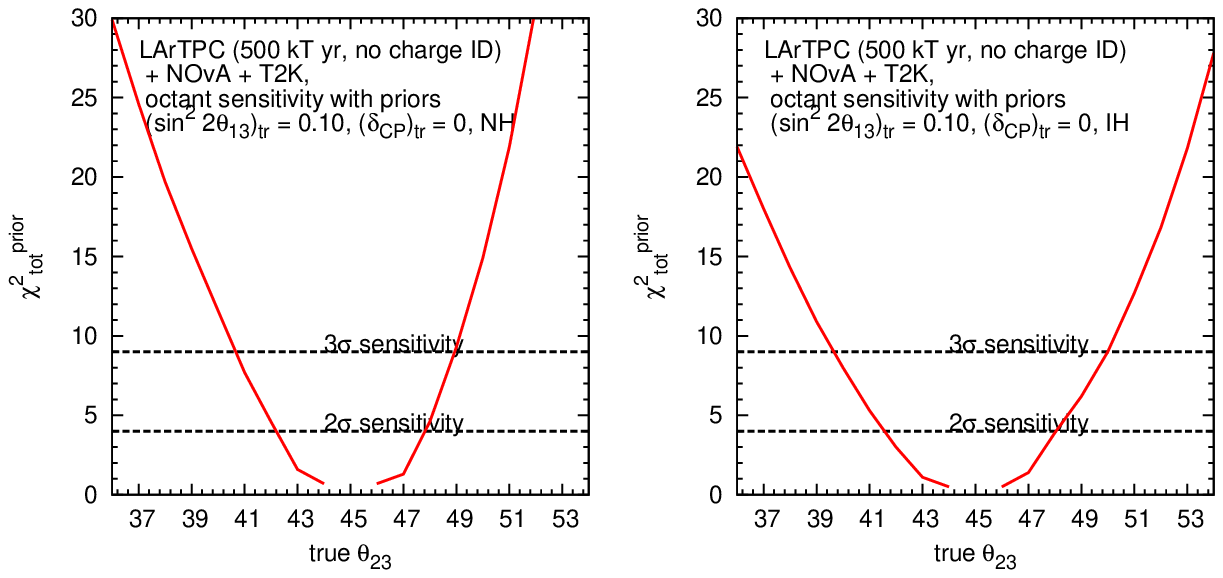}
\vspace{-0.4in}
\caption{\small Marginalized octant sensitivity from a combination of the atmospheric 
muon $+$ electron neutrino signal in a LArTPC (500 kT yr, without 
charge ID) $+$ \nova\ $+$ T2K, 
for the case of normal and inverted mass hierarchy. In this figure 
priors have been added.}
\label{octant_LArNovaT2K}
}

\begin{table}
%\TABLE[!h]{\centerline{
\begin{center}
%\TABULAR
\begin{tabular}{|c || c | c | c | c |} \hline
        {$\theta_{23}^{tr}$} &  LArTPC
& Mag.Iron &  LArTPC+\nova+T2K & Mag.Iron+\nova+T2K
        \\
        \hline
    36 & 19.8 (11.5) & 8.5 (3.9) & 29.9 (21.9) & 21.1 (19.9) \\ \hline
     37 & 15.6 (9.5) & 6.8 (3.2) & 24.6 (18.0) & 17.0 (16.3) \\ \hline
     38 & 12.4 (7.0) & 5.3 (2.5) & 19.7 (14.3) & 13.2 (12.9) \\ \hline
        39  & 8.9 (5.5) & 3.9 (1.9) & 15.5 (10.9) & 9.1 (9.8) \\ \hline
         40 & 5.9 (3.5) & 2.7 (1.3) & 11.6 (8.0) & 7.1 (7.1) \\ \hline
  41 & 4.0 (2.5) & 1.5 (0.9) & 7.7 (5.3) & 4.9 (4.8)   \\ \hline
  42 & 2.5 (1.6) & 0.6 (0.5) & 4.6 (3.0) & 3.2  (2.6) \\ \hline
  43 & 1.5 (0.8) & 0.2 (0.2) & 1.6 (1.1) & 1.4 (1.0)  \\ \hline
     44  & 0.6 (0.6) & 0.1 (0.1) & 0.7 (0.5) & 0.2 (0.2) \\ \hline
%        0.12  & 15.5 &  8.3 & 12.1
%\\ \hline
%        0.15  & 6.1 & 7.1 & 11.8 & 16.9

%       $\sin^2 \theta_{23} $ & 0.50 &  0.34 -- 0.68 \\ \hline
%       $\sin^2 \theta_{13} $ & 0.00 &  $\le$ 0.047
           \hline
\end{tabular}
\caption{\small Marginalized octant sensitivity with a LArTPC
(500 kT yr exposure), a magnetized iron detector (500 kT yr exposure), and a
combination of either of them with \nova\ $+$ T2K, for the case of normal
(inverted) mass hierarchy, with $\sin^2 2\theta_{13}^{tr} = 0.1$ and
$\dcp^{tr} = 0$. Priors have been included in this table.}
\label{chisqfinal}
\end{center}
\end{table}

In this section we present the octant sensitivity from a combined
analysis of simulated  long baseline and atmospheric data. 
We add ${\chi^2}_{prior}$ to the combination to take into account the 
future precision measurements on $\theta_{13}$ 
and $\theta_{\mu \mu}$ and minimize the ${\rm{(\chi^2_{tot})^{prior}}}$ given 
as,  

\bea
%{\rm{
%\chi^2_{tot}}} = {\rm{min(\chi^2_{NOvA} + \chi^2_{T2K} + \chi^2_{Atm})}} \nonumber \\
{\rm{(\chi^2_{tot})^{prior}}} = {\rm{min(\chi^2_{NOvA} + \chi^2_{T2K} + \chi^2_{Atm} + \chi^2_{prior})
}}
\eea
 Figure~\ref{octant_ICALNovaT2K} shows the combined octant sensitivity 
using the atmospheric muon neutrino signal in a magnetized iron detector with \nova\ and 
T2K. 
In Figure~\ref{octant_LArNovaT2K} the octant sensitivity computed from a 
combination of the atmospheric electron and muon neutrino signal 
in a LArTPC without charge identification capability with 
\nova\ and T2K is plotted.
%The combined analysis of atmospheric and fixed baseline experiments 
%is seen to improve the sensitivity significantly. 
In in Table~\ref{chisqfinal} we list the marginalized octant sensitivity 
with priors for a LArTPC (500 kT yr exposure), a magnetized iron detector 
(500 kT yr exposure), and 
a combination of each of them with \nova\ $+$ T2K, for the case 
of both NH and IH. All of the above results are for $\dcp^{tr}=0$.
For a magnetized iron detector, the combination with \nova\ $+$ T2K gives a 3$\sigma$ sensitivity  
at $\theta_{23} = 39^o$ for NH. A non-magnetized LArTPC $+$ \nova\ $+$ T2K gives a
4$\sigma$ sensitivity in the same case.
%Here the mass hierarchy is assumed to be known.
There is a tension between the behaviour of the \nova/T2K octant sensitivity and the 
octant sensitivity from an atmospheric neutrino experiment as a function 
of test $\theta_{23}$, which can be seen by comparing Figures~\ref{NovaT2K_testth23} 
and \ref{octant_testth23}. For \nova/T2K, the $\chi^2$ minima
occur at or close to $\theta_{23}^{test} = \pi/2 - \theta_{23}^{tr}$, while for an 
atmospheric experiment the muon events may have $\chi^2$ minima anywhere
between $\theta_{23}^{test} = 45^o$ and $\pi/2 - \theta_{23}^{tr}$ and the electron 
events have minima close to $\theta_{23}^{test} = 45^o$. 
This synergy leads to an enhancement of the octant sensitivity when the \nova\ $+$ 
T2K and atmospheric neutrino experiments are combined. 
%The results presented in Table \ref{chisqfinal} are for $\delta_{CP}^{tr}=0$. 

The combined $\chi^2$ values depend strongly on the true value 
of $\delta_{CP}$ following  the behaviour of the \nova/T2K $\chi^2$,
as can be seen from the green (dot-dashed) curve 
in Figure~\ref{octant_dcpt23}.
The LArTPC+\nova+T2K combination   
is seen to reach  a sensitivity of $> 4 \sigma$ for all
values of $\delta_{CP}$  for NH for 
the sample values of $\theta_{23}$  of $39^o$  and $51^o$ considered  
(the top and bottom left panels). 
For IH and $39^o$ ($51^o$)  one achieves close to 3$\sigma$ (4$\sigma$) 
sensitivity.
%For true HO and 
% has its minimum sensitivity
%at $\dcp^{tr} = 0$ in the case of true LO and NH (upper left panel) 
%and a sensitivity $> 4 \sigma$ is seen to be achievable for all 
%values of $\delta_{CP}$. 
The combination of the magnetized iron 
detector with \nova+T2K  is also expected to have a similar behaviour 
with $\delta_{CP}$ albeit with a lower sensitivity. 
The \nova\ and T2K contribution play a greater role in this case 
than that for a LArTPC. 
The sensitivity for \nova+T2K is higher for IH than for NH
for $\dcp^{tr}=0$ and $\theta_{23}^{tr}$ lying in the LO
(as seen in Table~\ref{tablelimits_NovaT2K}).
Therefore for these parameter values, 
the octant sensitivity for a magnetized iron detector+\nova+T2K combination
has comparable values for NH and IH, even though the sensitivity from a magnetized iron detector alone
is less for IH than for NH. In the case of a LArTPC+\nova+T2K combination, because of a greater contribution 
from LArTPC, the sensitivity is less for IH than for NH, reflecting 
the behaviour of the sensitivity from LArTPC.
This feature can be observed in Table~\ref{chisqfinal}.

{{Finally, we present the results of a combined analysis of an unmagnetized LArTPC with a magnetized iron detector in Table~\ref{chisqfinal2}, which lists
the octant sensitivities for a LArTPC+iron detector combination as well as a LArTPC+iron detector+\nova+T2K combination.
The results are seen to improve further in this case. The synergy
between these experiments helps in completely removing the effect of marginalizing over the hierarchy, since a magnetized iron detector itself gives good hierarchy discrimination
for the detector exposure considered, as discussed in Section 3.5. A combination of a magnetized iron detector with a LArTPC (500 kT yr)
and \nova + T2K gives a sensitivity of 4.5$\sigma$ for
$\theta_{23} = 39^o$ and nearly 6$\sigma$ for $\theta_{23} = 37^o$
for NH, and nearly 5$\sigma$ for $\theta_{23} = 37^o$ 
for IH. These results are unaffected by a marginalization over the hierarchy}}.

\begin{table}
%\TABLE[!h]{\centerline{
\begin{center}
%\TABULAR
\begin{tabular}{|c || c | c |} \hline
        {$\theta_{23}^{tr}$} & Mag.Iron+ LArTPC & Mag.Iron+LArTPC+\nova+T2K
        \\
        \hline
    36 & 29.1 (15.6) &  41.2 (27.0)  \\ \hline
     37 &  22.8 (12.9) & 33.9 (22.3)  \\ \hline
     38 &  18.5 (9.7) & 27.1 (17.7)  \\ \hline
        39  & 13.8 (7.6) & 21.0 (13.5) \\ \hline
         40 & 9.1 (5.0) & 15.4 (9.8)  \\ \hline
  41 & 6.4 (3.5) & 10.1 (6.5)  \\ \hline
  42 &  3.1 (2.1) & 5.9 (3.7) \\ \hline
  43 &  1.7 (1.0) & 1.9 (1.4)  \\ \hline
     44  & 0.7 (0.7) & 0.9 (0.7) \\ \hline
%        0.12  & 15.5 &  8.3 & 12.1
%\\ \hline
%        0.15  & 6.1 & 7.1 & 11.8 & 16.9

%       $\sin^2 \theta_{23} $ & 0.50 &  0.34 -- 0.68 \\ \hline
%       $\sin^2 \theta_{13} $ & 0.00 &  $\le$ 0.047
           \hline
\end{tabular}
\caption{\small Marginalized octant sensitivity with a a combination of a LArTPC
(500 kT yr exposure) and a magnetized iron detector (500 kT yr exposure), as well as a combination of
both of them with \nova\ $+$ T2K, for the case of normal
(inverted) mass hierarchy, with $\sin^2 2\theta_{13}^{tr} = 0.1$ and
$\dcp^{tr} = 0$. Priors have been included in this table.}
\label{chisqfinal2}
\end{center}
\end{table}

\section{Summary and Conclusion} 

In this paper we have studied the  possibility of determining the 
octant of the atmospheric mixing angle $\theta_{23}$ in the long baseline 
experiments T2K and \nova\ as well as by atmospheric neutrino experiments. 
While the octant degeneracy conventionally refers to the indistinguishability
between $\theta_{23}$ and $\pi/2 - \theta_{23}$, this can be generalized to
include the whole range of allowed value of $\theta_{23}$ in the wrong octant,
and we consider this generalized definition in our analysis.

We present a probability level discussion on the 
effect of  uncertainty in $\theta_{13}$, $\delta_{CP}$
and values of  $\theta_{23}$
in the wrong octant for baselines relevant to long baseline and atmospheric
experiments.
Below we summarize the salient features that emerge from our study at the
probability level: 
 
\begin{itemize}

\item 
For baselines where matter effects are small,
the appearance channel probability $\pmue$ displays a
degeneracy of the $\theta_{23}$ octant with the values of $\theta_{13}$ and
$\dcp$
due to its dependence on the combination 
$\sin^2 \theta_{23} \sin^2 2\theta_{13}$
at leading order as well as its subleading $\dcp$ dependence.
The disappearance
channel probability $\pmumu$ suffers from an intrinsic octant
degeneracy between $\theta_{23}$
and $\pi/2 - \theta_{23}$ due to being a function
of $\sin^2 2\theta_{23}$ at leading order.

\item For \nova/T2K baselines, there is a strong effect of the uncertainty
in $\dcp$
in the appearance channel, and of the $\theta_{23}$ uncertainty in the  
disappearance
channel. 
% For a given value of $\theta_{13}$, the lack of knowledge of $\delta_{CP}$
%can give rise to a degeneracy between the two octants in $\pmue$. 
For instance, for
the \nova\ baseline
the degeneracy between $\theta_{23}=39^o$ and $51^o$ in $\pmue$ is lifted only
for $\sin^2 2\theta_{13} \gsim 0.12$ if a variation over the entire range of $\dcp$
is taken into account. 

%The $\pmumu$ channel is less affected by $\dcp$ but shows a
%significant effect of the intrinsic degeneracy. The effect of these degeneracies is
%more pronounced for shorter baselines like these, where matter effects are very small.

\item 
%Hence a high value and good precision of $\theta_{13}$ is required in order to be able
%to exclude degenerate solutions in the wrong octant. 
We find that after including the
improved precision in $\theta_{13}$
from the recent reactor results, the octant degeneracy with respect to $\theta_{13}$
is largely reduced. Some amount of degeneracy with respect to $\theta_{13}$
still remains for higher (lower) values of $\theta_{13}$ in the allowed range
for a true lower (higher) octant. Also, in general, information from the energy 
spectrum helps in reducing the degeneracy since for two different energies the
degeneracy exists for different sets of parameter values.

\item For a sample baseline of 5000 km traversed by the 
atmospheric neutrinos 
% like those for atmospheric neutrinos (2000 - 12000 Km),
strong resonant matter effects help in lifting both the intrinsic octant degeneracy
in $\pmumu$ and the  octant degeneracy with $\theta_{13}$ and $\dcp$ in $\pmue$.
This is because the term $\sin^2 2\theta_{13}^{\rm{m}}$ becomes close to 1 at or
near matter resonance, and this makes the leading-order term proportional to
$\sin^2 \theta_{23}$ ($\sin^4 \theta_{23}$) in $\pmue$ ($\pmumu$) predominate and
give distinct values in the two octants  irrespective of the vacuum value of
$\theta_{13}$. 

%The effect is large enough to overrule any corresponding enhancement
%in the sub-leading $\dcp$-dependent terms. In $\pmumu$, this behaviour overrides
%the effect of the intrinsic octant degeneracy arising from the $\sin^2 2\theta_{23}$
%dependence. Thus the degeneracy is resolved with respect to $\theta_{13}$ and
%$\delta_{CP}$ as well as different values of $\theta_{23}$ in the wrong octant

\end{itemize} 

We perform a $\chi^2$ analysis of the octant sensitivity for 
T2K/\nova\ and atmospheric neutrinos as well as a combined study. 
For atmospheric neutrinos  we consider two types of detectors -- 
magnetized iron calorimeter detectors 
capable of detecting muons and identifying their charge and a LArTPC 
(non-magnetized) which  can detect both muons and electrons with superior 
energy and angular resolutions.
%Since magnetized iron calorimeter detectors
%have very good sensitivity to
%the hierarchy, it is assumed that the hierarchy will be determined by
%these detectors
%before the octant. 
For \nova/T2K we present the results for both cases of known and unknown hierarchy 
while the atmospheric results are assuming hierarchy to be known. 
The main results are summarized below: 

\begin{itemize} 

\item The $\chi^2$ minima for the \nova/T2K disappearance channel occur near
$\theta_{23}^{test} = \pi/2 - \theta_{23}^{tr}$, because of the predominant
dependence on $\sin^2 2\theta_{23}$, while for the appearance channel the minima
occur near $\theta_{23}^{test} = 45^o$, because of the $\sin^2 \theta_{23}$ dependence.
This leads to a tension between the $\chi^2$ behaviour of the two spectra 
as a
function of $\theta_{23}^{test}$, so that the combination of appearance and
disappearance channels enhances octant sensitivity,  but even then 
the $\chi^2$ is not very high. 
Combining the \nova\ and T2K data leads to a higher octant sensitivity due to the
addition of sensitivities from the two experiments.
What plays a major role in enhancing the octant sensitivity in 
these experiments is the 
addition of priors, especially
on $\theta_{13}$, which helps in ruling out the degenerate solutions 
in the wrong octant. After adding priors one can achieve a $\sim 2 \sigma$ 
sensitivity  at $\theta_{23} = 39^o$ for $\sin^2 2\theta_{13}=0.1$
and $\delta_{CP}=0$ for both normal and inverted hierarchies.

\item
A magnetized iron calorimeter gives a 2$\sigma$ sensitivity to the octant for
$\theta_{23}= 39^o$ and $\sin^2 2\theta_{13}=0.1$ for a true normal hierarchy.
A non-magnetized LArTPC can give a 3$\sigma$ sensitivity  for the 
same parameter values.
The enhanced sensitivity of the LArTPC is due to the contribution of the
electron events as well as superior energy and angular resolutions compared 
to an iron
calorimeter detector. In a LArTPC, 
combining the muon and electron events gives improved
sensitivities
due to the different behaviour of the muon and electron event spectra with respect to 
$\theta_{23}$. 
If a LArTPC can have charge identification capability there will be a
20-40$\%$ 
increase in sensitivity, the enhancement being more in the case of an 
inverted hierarchy than for a normal hierarchy.
Since we assume hierarchy to be already known, the magnetization of 
LArTPC (which is technically challenging) does not play a significant role. 
{{A marginalization over the hierarchy
does not affect the results for a magnetized iron detector, since
it already excludes the wrong hierarchy with a good confidence level
for the same exposure. For a LArTPC, the results are unchanged 
by a hierarchy marginalization over the range $39^o \leq \theta_{23} \leq 51^o$, but
suffer a drop of 20-30$\%$ for values of $\theta_{23}$ above and below this}}. 

\item Combining long baseline and atmospheric results can give a significant
enhancement in the octant sensitivity because of the tension in the behaviour of
the \nova/T2K and atmospheric $\chi^2$ as functions of test $\theta_{23}$,
arising
due to the different energy and baseline ranges involved and the fact that 
strong
matter effects dictate the probability behaviour for atmospheric baselines. 
Since
the $\chi^2$ minima occur at different parameter values, there is a synergy between
the experiments which gives an increased octant sensitivity. A combination of
\nova, T2K and LArTPC (500 kT yr) can give a nearly 3$\sigma$ sensitivity
for $\theta_{23} = 41^o$, a 4$\sigma$ sensitivity for $\theta_{23} = 39^o$
and a 5$\sigma$ sensitivity for $\theta_{23} = 37^o$ for $\sin^2 2\theta_{13} = 0.1$
and $\dcp = 0$ in the case of a normal hierarchy.  The corresponding 
sensitivities for inverted hierarchy are somewhat less.  
A magnetized iron calorimeter combined with \nova\ and T2K
can give $\sim 3\sigma$ sensitivity for $\theta_{23} = 39^o$ 
for both NH and IH.  
{{Finally, a combination of a magnetized iron detector with a LArTPC (500 kT yr)
and \nova + T2K gives a further improved sensitivity of 4.5$\sigma$ for
$\theta_{23} = 39^o$ and nearly 6$\sigma$ for $\theta_{23} = 37^o$
for NH, and nearly 5$\sigma$ for $\theta_{23} = 37^o$ 
for IH. These results hold good with a marginalization over the hierarchy}}.

\item The octant sensitivity from an atmospheric neutrino experiment
is nearly independent of the true value of $\dcp$,  while the \nova\ and T2K sensitivities
are seen to be  strongly $\dcp$-dependent and follow a definite shape as
a function of $\dcp$. The shape gets flipped  for a true lower or higher
octant. For a normal mass hierarchy, the contribution of \nova\ $+$ T2K to the octant
sensitivity is lower than that from a LArTPC measurement for almost all values
of $\dcp$, while for an inverted hierarchy it is  higher, due
to the worse sensitivity from an atmospheric experiment in the inverted hierarchy case.

\end{itemize}

In conclusion, we find that the improved precision of $\theta_{13}$ 
and the different dependence on $\theta_{23}$  in the disappearance 
and appearance channels and in vacuum and matter probabilities 
leads to an enhanced octant sensitivity  when long baseline and 
atmospheric neutrino experiments are combined.  

%\section{Points to be checked after writing the whole paper}
%\begin{itemize}
% \item Have we correctly used $\theta_{23}$ and $\theta_{\mu\mu}$ everywhere, depending on 
% context?
% \item NOvA or \nova
% \item Have we avoided using the terms INO and ICAL everywhere?
% \item kt or kton or kiloton? Mt or Mton or megaton?
% \item Liquid Argon or Liq Ar or LArTPC or ...?
% \item Placement of figures and tables vis \`{a} vis their reference in the text
% \item Figures spilling out of textwidth
% \item Figures/Tables covering up the line just above them
%\end{itemize}

\bibliographystyle{JHEP}
\bibliography{octant2012}

\end{document}